# Understanding the Role of Triplet-triplet Annihilation in Non-fullerene Acceptor Organic Solar Cells

Lucy J. F. Hart[1,2], Jeannine Grüne[1,3], Wei Liu[4], Tsz-ki Lau[5], Joel Luke[6], Yi-Chun Chin[6], Xinyu Jiang[7], Huotian Zhang[8], Daniel J.C. Sowood[1], Darcy M. L. Unson[1], Ji-Seon Kim[6], Xinhui Lu[5], Yingping Zou[4], Feng Gao[8], Andreas Sperlich[3], Vladimir Dyakonov[3], Jun Yuan[4*] and Alexander J. Gillett[1*].

[1]Cavendish Laboratory, University of Cambridge, JJ Thomson Avenue, Cambridge, U.K.
[2]Department of Chemistry and Centre for Processable Electronics, Imperial College London, 82 Wood Lane, London, U.K.
[3]Experimental Physics 6, Julius Maximilian University of Würzburg, Am Hubland, 97074 Würzburg, Würzburg, Germany
[4]College of Chemistry and Chemical Engineering, Central South University, Changsha 410083, P.R. China.
[5]Department of Physics, The Chinese University of Hong Kong, Shatin, 999077 Hong Kong.
[6]Department of Physics and Centre for Processable Electronics, Imperial College London, South Kensington, U.K.
[7]Chair for Functional Materials, Department of Physics, TUM School of Natural Sciences, Technical University of Munich, James-Franck-Str. 1, 85748 Garching, Germany.
[8]Department of Physics, Chemistry and Biology (IFM), Linköping University, Linköping, Sweden.

*Corresponding authors: Alexander J. Gillett: E-mail: ajg216@cam.ac.uk; Jun Yuan: E-mail: junyuan@csu.edu.cn.

**Abstract**

Non-fullerene acceptors (NFAs) have enabled power conversion efficiencies exceeding 19% in organic solar cells (OSCs). However, the open-circuit voltage of OSCs remains low relative to their optical gap due to excessive non-radiative recombination, and this now limits performance. Here, we consider an important aspect of OSC design, namely management of the triplet exciton population formed after non-geminate charge recombination. By comparing the blends PM6:Y11 and PM6:Y6, we show that the greater crystallinity of the NFA domains in PM6:Y11 leads to a higher rate of triplet-triplet annihilation (TTA). We attribute this to the four times larger ground state dipole moment of Y11 versus Y6, which improves the long range NFA out-of-plane ordering. Since TTA converts a fraction of the non-emissive triplet states into bright singlet states, it has the potential to reduce non-radiative voltage losses. Through a kinetic analysis of the recombination processes under 1-Sun illumination, we provide a framework for determining the conditions under which TTA may improve OSC performance. If these could be satisfied, TTA has the potential to reduce non-radiative voltage losses by up to several tens of mV and could thus improve OSC performance.

**Main Text**

*1. Introduction*

The past five years have seen a rapid improvement in organic solar cells (OSCs), with record device efficiencies in single junctions jumping from 12% to over 19% [1,2]. Much of this improvement can be ascribed to the development of efficient non-fullerene electron acceptor (NFA) materials, also known as small molecule acceptors (SMAs) [3,4]. One prominent family of narrow bandgap SMAs is the 'Y-series', two examples being Y6 and Y11. These NFAs are typically combined with the donor material PM6 to create PM6:Y6 and PM6:Y11 bulk heterojunctions (molecular structures shown in Figure 1a), which have formed the basis for OSCs with efficiencies > 16% [5,6]. However, even the most efficient OSCs still have an open-circuit voltage ($V_{OC}$) significantly below the radiative limit. This discrepancy is attributed to non-radiative voltage losses ($\Delta V_{nr}$), which are higher in OSCs than in their inorganic counterparts. The magnitude of $\Delta V_{nr}$ can be obtained from the electroluminescence external quantum efficiency ($EQE_{EL}$) of the device run at forward bias using Rau's reciprocity relationship, $\Delta V_{nr} = - k_B T \ln(EQE_{EL})$ [7]. To further aid understanding, it is helpful to separate the $EQE_{EL}$ into four components:

$$EQE_{EL} = \gamma \varphi_{PL} \chi \eta_{OUT} \tag{1}$$

$\gamma$ is the charge balance factor, $\varphi_{PL}$ is the photoluminescence quantum yield (PLQY), $\chi$ is the fraction of recombination events which can occur radiatively, and $\eta_{OUT}$ is the photon out-coupling efficiency. Whilst the OSC community has focused on improving the PLQY, the contribution from $\chi$ to $\Delta V_{nr}$ has received relatively little attention until recently. As is already well known for organic light-emitting diodes (OLEDs), the value of $\chi$ in OSCs will generally be less than 1 due to the spin-statistics of free charge recombination, which predict a 75% yield of spin-triplet states [8]. Following the formation of the molecular triplet exciton on the low band gap blend component, decay back to the ground state will generally proceed non-radiatively via triplet-charge annihilation (TCA) or monomolecular triplet decay, leading to an increase in $\Delta V_{nr}$ by up to 60 mV [9-12]. Thus, finding ways to manage the triplet population is now crucial for further improving the $V_{OC}$ values of NFA-based OSCs.

In this work, we examine the potential of triplet-triplet annihilation (TTA) to recycle the triplet excitons formed after non-geminate charge recombination in OSCs. In TTA, the interaction of two triplets can lead to one being excited to the singlet state and the other returning to the ground state (alongside other possible decay routes, see ref. 13 for a more detailed discussion), as shown schematically in Figure 1b. TTA has been extensively studied in the OLED field as it can increase $\chi$ from 0.25 to 0.625 and thus improve $EQE_{EL}$ [14]. Indeed, TTA is currently the mechanism employed in most commercial blue OLEDs [15]. By contrast, the potential of TTA to improve the $EQE_{EL}$ (and thus $V_{OC}$) of OSCs has not yet been considered, although there have been some reports of TTA in fullerene acceptor OSCs [16,17]. Through studying the contrasting behaviour of the triplet population in two high performance NFA OSC blends,

PM6:Y11 and PM6:Y6, we find that they differ with respect to the dominant triplet decay mechanism. While we confirm that TCA is the primary triplet decay mechanism in PM6:Y6 [9], our results suggest that triplet decay in PM6:Y11 is dominated by TTA. As such, we demonstrate that TTA is a possible triplet decay pathway in NFAs and, inspired by triplet management strategies in OLEDs, we proceed to investigate the feasibility of reducing $\Delta V_{nr}$ in OSCs using TTA.

*2. Results and Discussion*

To begin, conventional architecture PM6:Y11 and PM6:Y6 devices have been fabricated. Both devices show good performance with PCEs >15%, in line with previous reports [5,6] (see Figure S1 for device JV curves and the photovoltaic external quantum efficiency, $EQE_{PV}$). Thus, these are high performance material systems which are of relevance to current OSC research.

*2.1. Transient Absorption Spectroscopy*

To better understand the dynamics of the excited states in PM6:Y11 and PM6:Y6, we turn to femtosecond transient absorption spectroscopy (TAS). Following selective excitation of the NFA at 800 nm (see Figure S2 for the absorption spectra of both the neat NFAs and their blends with PM6), we present the results for the infrared (IR) spectral region (1200-1650 nm) in Figure 2a (PM6:Y11) and Figure 2b (PM6:Y6). The other spectral regions are shown in Figure S3. There are two distinct features in both IR-region spectra. Initially, there is a photoinduced absorption (PIA) peaking in the 1500-1600 nm region, which has previously been assigned in Y6 to an intermolecular CT-type state between neighbouring molecules (hereafter, an inter-CT state [18]) and is also seen in the TA spectrum of the neat films (Figure S4). We note that Y6 has a higher formation yield of inter-CT states from singlet excitons than Y11 and thus, for a given excitation intensity, the inter-CT signal will be relatively larger in Y6 (Figure S4). The inter-CT PIA decays within the first 100 ps of the measurement (Figure S3) as holes are transferred from the NFA to PM6 (Figure S5). After hole transfer is completed, there is a rise in a second PIA centred between 1400-1450 nm (see Figures 2a and 2b). In both blends, we assign this PIA to the NFA triplet exciton, based on the results of previous triplet sensitisation experiments on Y6 [9].

Following the identification of the excited species, their kinetics can be calculated from the time dependence of $\Delta T/T$ using the equation:

$$\frac{\Delta T(\lambda,t)}{T} = -w\sigma(\lambda)\Delta n(\lambda, t) \qquad (2)$$

$w$ is the film thickness, $\sigma(\lambda)$ is the absorption cross-section and $\Delta n(\lambda,t)$ is the density of the excited species (averaged over the film thickness). Figure 2c and Figure 2d show the time-dependent TA kinetics from the wavelength region associated with the Y11 and Y6 triplet exciton, respectively, over a series of fluences, normalised to the inter-CT state PIA around time zero. Although the signal due to the triplet exciton is convoluted with that of the inter-CT state at early times, the signal at later times can be ascribed solely to the NFA triplet exciton as hole transfer to the PM6 has been completed, quenching the inter-CT state PIA (see Figure S5). Thus, the fluence dependence of the growth of the triplet-region kinetics indicates that triplet formation is caused by a bi-molecular process, namely the non-geminate recombination of free charge carriers [9].

It is also notable that the triplet exciton population in PM6:Y11 starts to decay at a lower normalised $\Delta T/T$ value as fluence is increased. This behaviour is especially striking when the triplet kinetics of PM6:Y11 are compared to those of PM6:Y6, which display the opposite trend. At first, the behaviour of PM6:Y11 seems counter-intuitive; the increased charge carrier density at higher fluences results in more recombination events, meaning that the triplet population should reach an earlier maximum that is higher relative to the point of normalisation. Indeed, this is the behaviour observed in PM6:Y6. However, the extremely rapid and strongly fluence-dependent triplet quenching in PM6:Y11 implies that the dominant triplet decay mechanism has a higher order dependence on the triplet population than the TCA process previously reported to dominate in PM6:Y6 [9].

To better understand the mechanisms driving triplet decay in these blends, we modelled the data with the following rate equation:

$$\frac{dn_T}{dt} = -\alpha \frac{dn_H}{dt} - \beta n_P n_T - \gamma n_T^2 \qquad (3)$$

$n_T$ and $n_H$ refer to the triplet and hole population densities, respectively, $\alpha$ is the fraction of non-geminate recombination which leads to triplet formation, and $\beta$ and $\gamma$ are the rate constants of TCA and TTA, respectively. We note that this equation is a combination of previous models used to extract information about TCA [10] and TTA [16] processes. We exclude mono-molecular triplet decay from the rate equation as fits to nanosecond TAS data (discussed further in S1.2) indicate that the triplet lifetime in the neat NFA films (tens to hundreds of nanoseconds) exceeds the time scales of the femtosecond TAS data discussed presently. Additionally, although the neat NFAs have a non-negligible triplet population generated by intersystem crossing (Figure S6), this term is also neglected since the efficient charge transfer from the NFAs to the PM6 in the blend [19,20] outcompetes non-radiative NFA singlet exciton decay pathways. Thus, we only consider triplet excitons formed via non-geminate recombination.

Fitting equation 3 to the data directly requires knowledge of the absorption cross-sections for the PM6 hole polaron and the NFA triplet excitons. Due to the difficulty of extracting accurate values for these quantities from the femtosecond TAS data, we here report the results of fitting the 'cross-section free' equation:

$$\frac{d}{dt}\left(\frac{\Delta T}{T}\bigg|_T\right) = -A\frac{d}{dt}\left(\frac{\Delta T}{T}\bigg|_P\right) - B\left(\frac{\Delta T}{T}\bigg|_P\right)\left(\frac{\Delta T}{T}\bigg|_T\right) - G\left(\frac{\Delta T}{T}\bigg|_T\right)^2 \qquad (4)$$

where the T (triplet) and P (polaron) subscripts denote the species responsible for the measured ΔT/T and $A$, $B$ and $G$ are fitting parameters proportional to $\alpha$, $\beta$ and $\gamma$, respectively (see S1.1 for a full description of the fitting procedure). The results of the fits for PM6:Y11 and PM6:Y6 are shown in Figure 2e and Figure 2f, respectively and the fitting parameters are given in Table 1. Although this method of fitting does not allow for a quantitative comparison between the two blends, it still allows us to determine the origin of the differing triplet dynamics in PM6:Y6 and PM6:Y11. In PM6:Y6, values for both the TCA ($B = (6.3 \pm 0.6) \times 10^{10}$ s$^{-1}$) and TTA ($G = (7 \pm 2) \times 10^{10}$ s$^{-1}$) rate constants could be extracted, which were of comparable size. As in demonstrated in S3.2, the polaron population in PM6:Y6 is significantly larger than the triplet population meaning that the dominant triplet decay pathway in PM6:Y6 is TCA. By contrast, in PM6:Y11, we could only extract a value for $G$ ($G = (2.3 \pm 0.2) \times 10^{11}$ s$^{-1}$), not $B$, which indicates that there is negligible TCA in this blend and that triplet decay is instead dominated by TTA (see Figures S7-9 for further discussion of this point).

To investigate if the rate of TTA is comparable between the neat NFAs and their blends with PM6, we performed nanosecond TAS on neat films to extract their TTA rates, as detailed in S1.2. The fit results are shown in Figure S10, and the parameters are summarised in Table S1. These indicate that the $G$ value is $(1.46 \pm 0.04) \times 10^{11}$ s$^{-1}$ in neat Y11 and $(1.20 \pm 0.04) \times 10^{11}$ s$^{-1}$ in neat Y6. As described in S1.1, the rate of TTA ($\gamma$) depends on the film thickness as well as $G$. Thus, to compare these $G$ values to those of the neat materials, we need to account for the fact that the neat films are approximately half the thickness of the blend films and so we expect $G$ to be double the size in the neat film for the same rate of TTA (i.e., the same $\gamma$). While this is the case when comparing the $G$ values of neat Y6 and PM6:Y6, the $G$ value in neat Y11 is in fact smaller than that in PM6:Y11, implying that the blend film has a significantly enhanced rate of TTA when compared to the neat film. To explain this discrepancy, we note that the neat Y11 film was not thermally annealed, unlike the PM6:Y11 blend. As is shown by the GIWAXS data discussed below, thermal annealing significantly improves the molecular ordering of PM6:Y11, which may affect the triplet mobility and thus the rate of TTA, as TTA is often considered to be a diffusion limited process [21-24]. To investigate this hypothesis, we performed femtosecond TAS on an unannealed PM6:Y11 sample and fitted the data using the same method as above to extract the rate of TTA (Figure S11). This was found to be $G = (1.7 \pm 0.3) \times 10^{11}$ s$^{-1}$, which is lower than that of the thermally annealed film, thereby supporting the hypothesis that improved NFA crystallinity increases the rate of TTA.

*2.2. Photoluminescence Detected Magnetic Resonance*

As illustrated schematically in Figure 1b, TTA can result in the formation of singlet excitons which are then able to decay radiatively. Thus, TTA allows 'dark' triplet states to contribute indirectly to the total photoluminescence (PL), allowing them to be detected using optical methods [25,26]. To probe this, we use photoluminescence detected magnetic resonance (PLDMR), a spin-sensitive PL technique which can be used to detect triplet states that are coupled to luminescence, e.g. via TTA, ground state depletion, or (reverse) intersystem crossing [27]. Furthermore, since PLDMR uses continuous wave (cw) illumination, it provides a better approximation of the conditions in real devices where the accumulation of triplet excitons can lead to an increased probability of annihilation effects, including TTA [28,29].

Figure 3a shows the cwPLDMR spectra of PM6:Y11 and PM6:Y6, while those of the neat materials (PM6, Y11 and Y6) are shown in Figure S12. The full-field (FF) spectrum (280 – 420 mT) corresponds to $\Delta m_S = \pm 1$ transitions between triplet sublevels. The width of this signal is a measure of the axial zero-field splitting (ZFS) parameter $D$, which is correlated to the interspin distance $r$ ($D \sim r^{-3}$) [30,31]. Thus, the middle, narrow peak ($B = 336.1$ mT) corresponds to distant spin centers, such as CT states, while the broad signal is associated with molecular triplet excitons. The ZFS parameters of the broad PLDMR feature are found to be $D = 930$ MHz and $E = 140$ MHz for PM6:Y11 and $D = 990$ MHz and $E = 140$ MHz for PM6:Y6 (EasySpin simulations and parameters are shown in Figure S13 and Table S2 in the SI). Furthermore, both blends show a half field (HF) signal at $B = 167$ mT, corresponding to first-order forbidden $\Delta m_S = \pm 2$ transitions between T$_+$ and T$_-$ sublevels. As the parameters of the FF and HF signals in the blend films are consistent with the PLDMR of the neat NFAs, this confirms the presence of the NFA molecular triplet species in both blends.

The shape of the PLDMR spectra depends on the molecular orientation relative to the external applied magnetic field, which is given by the angle $\theta$ (see inset of Figure 3a). For $\theta = 0°$, the PLDMR spectra show a clear preferential orientation of the molecules, indicated by the 'wings' of the spectrum [31]. This axial alignment can be described by an ordering factor $\lambda_\theta$, weighting the anisotropy of the orientation distribution of the paramagnetic molecules [32,33]. The observed preferential alignment is consistent with reports in the literature, which demonstrate particularly pronounced intermolecular and substrate face-on stacking for Y-series NFAs [34-37]. Orientation dependent PLDMR measurements (Figure S14 and S2.2) show that the PLDMR wings at 0° are determined by the ZFS tensor component along the molecular z-axis. At 0°, the molecular z-axis is aligned with the external magnetic field, which corresponds to the out-of-plane (OOP) direction [32]. While PLDMR of the neat NFAs show a comparable ordering (Figure S12 and Table S2), the preferential orientation is increased in PM6:Y11 ($\lambda_\theta = 9.0$) in comparison to PM6:Y6 ($\lambda_\theta = 5.5$), as represented by the steeper wings in PM6:Y11 (Figure 3a). This enhanced alignment in PM6:Y11 suggests a higher OOP crystallinity of PM6:Y11 than in PM6:Y6, in agreement with the GIWAXS results discussed below.

To determine the process by which triplet excitons on Y11 couple to the PL, we performed laser power dependent transient PLDMR (trPLDMR). Figure 3b shows PLDMR transients for PM6:Y11, measured at $B = 304.5$ mT (the most pronounced triplet signal as measured via cwPLDMR) and $\theta = 0°$. The PLDMR signal is positive ($\Delta PL/PL > 0$), corresponding to an increase in PL under resonant conditions. This effect can arise from PL enhancement due to TTA and we proceed to confirm this by studying the behavior of the spectra as the laser excitation power is varied [38,39].

The amplitudes of the PLDMR transients for different laser excitation powers are fitted using the power law

$$\frac{\Delta PL}{PL} \sim \frac{P_{exc}^a}{P_{exc}^b} = P_{exc}^c \quad (5)$$

with $a$, $b$ and $c$ describing the power dependencies of the trPLDMR signal ($\Delta PL$), the total PL and the relative trPLDMR signal ($\Delta PL/PL$), respectively [38,40,41]. In Figure 3c we show the result for $\Delta PL \sim P_{exc}^a$ to directly evaluate the power dependence of the triplet-sensitive $\Delta PL$ signal, while the results for $\Delta PL/PL \sim P_{exc}^c$ are given in Figure S15. In both cases, there is a clear division of the data into a low-power regime (≲ 10 mW) and a high-power regime. In the low-power regime, the fit of the power law gives a slope of $a = 1.47 \pm 0.05$, while this decreases to $a = 1.00 \pm 0.04$ in the high-power regime. Conventional TTA upconversion systems show a quadratic increase in PL at lower excitation intensities, which transitions to a linear increase at higher excitation intensities, once a certain threshold intensity, $I_{th}$, is crossed [41-45]. The reason for this transition is that the dominant decay pathway of the triplets shifts from being a monomolecular process at lower intensities to TTA at high intensities. When TTA becomes the main triplet decay pathway, the upconversion yield reaches its maximum, resulting in only a linear increase with excitation density, also called "annihilation-limited" regime [41,42,46].

The slope in the low-power regime ($a = 1.47$) deviates from the value of $a = 2$ measured in conventional TTA upconversion systems. However, Izawa et al. have investigated the energy transfer from Y6 to the TTA material rubrene and reported similar power dependences for the PL: 1.57 in the low-power regime and 1.00 in the high-power regime [46]. This deviation from the conventional behaviour may be due to contributions from other, bimolecular processes, such as TCA or singlet-singlet annihilation (see Figure S6 for the latter) [46]. Thus, we conclude that the triplet-sensitive PL (ΔPL) presented here shows power law behaviour which is typical of TTA systems. Importantly, the detection of two power regimes allows us to confirm that TTA is the dominant decay channel for triplet excitons in PM6:Y11 at higher excitation powers.

Another indication of increased TTA involvement in PM6:Y11 can be obtained by comparing the ratio of the broad triplet exciton feature to the middle CT state peak (Figure S16). For this, we use trPLDMR as it is independent of modulation aspects which affect the cwPLDMR measurements. Although there are different factors which can influence the size of the middle peak (e.g. enhanced TCA), the ratio of the triplet-sensitive PL to the CT state middle peak is nine times larger in PM6:Y11 than in PM6:Y6 (0.45 versus 0.05). Since the TAS measurements indicate that both blends possess a similar triplet population for a given excitation fluence and it is assumed that the spin polarization of the triplet sublevels are comparable in both material blends due to the otherwise similar photophysical processes, we propose that there is a higher coupling constant of the triplet to the singlet state in PM6:Y11, i.e., a higher TTA rate. This may be caused by a different NFA stacking motif, as suggested by the increased ordering parameter ($\lambda_\theta$) observed in PM6:Y11.

*2.3. Grazing-Incidence Wide-Angle X-ray Scattering (GIWAXS)*

To gain additional confirmation of the enhanced ordering in PM6:Y11, we consider the GIWAXS of PM6:Y11, PM6:Y6, neat Y11, neat Y6 and neat PM6 films shown in Figure 4. The line-cuts for the neat materials along the in-plane (IP) and OOP directions are given in Figure 4a and Figure 4b, respectively, and the corresponding line cuts for the blend films are given in Figure 4c and Figure 4d. The *d*-spacings of the peaks are reported in Table S3 and the 2D GIWAXS images are shown in Figure S17. In PM6, the strong lamellar (100) peak at q~0.3 Å$^{-1}$ in both the IP and OOP directions suggests a relatively isotropic ordering of the polymer chains [47]. Moving to the NFA films, both have a pronounced peak in intensity at q~0.4 Å$^{-1}$ along the IP direction, with a strong (010) peak at q~1.7 Å$^{-1}$ in the OOP direction. This demonstrates that neat, unannealed Y11 and Y6 have a similar molecular orientation, which is strongly face-on to the substrate [37,48,49]. However, there are differences between the IP peaks of the two NFAs. In Y6, two peaks are visible with d-spacings of 21.9 Å and 14.9 Å, attributed to the two-dimensional structure order in the backbone plane [50], while, in Y11, only one peak can be discerned with a d-spacing of 15.3 Å. The absence of the peak around 21.9 Å in Y11 may indicate that Y11 favours Core-Terminal stacking, since the distance between the end groups of Y11 is ~ 22 Å, while the distance between end group and core group is ~ 15 Å [50]. The broader peak at q~1.7 Å$^{-1}$ in the OOP (010) direction is due to the π-π stacking of the NFA molecules. In Y11, this peak has a greater intensity and occurs at a lower $q_z$ value than is the case for Y6, indicating stronger π-π stacking with a slightly larger stacking distance in Y11. These differences in crystallinity could contribute to the different rates of TTA and triplet lifetimes [51] which were found in the nanosecond TAS measurements of the neat, unannealed NFAs (see S1.2 and Table S1).

In the PM6:Y11 and PM6:Y6 films, the peaks identified in the neat materials are still visible, though some peaks shift to lower q values (Table S3), which may indicate a slight enhancement in the component material's crystallinity in the blend film [52]. However, the most striking feature of the PM6:Y11 GIWAXS when compared to that of PM6:Y6 is the new scattering peak at q~0.6 Å$^{-1}$ along the OOP direction. We note that this feature only emerges in PM6:Y11 following annealing, which is why it is not present in the GIWAXS of the as-cast Y11 film [6]. Its presence suggests that PM6:Y11 has a greater degree of long-range ordering in the OOP direction than PM6:Y6, as was already suggested by the PLDMR (Figure 3a).

To better understand the origin of the enhanced ordering in PM6:Y11, the equilibrium geometry and dipole moment of Y6 and Y11 molecules were calculated using density functional theory (calculation details given in the Experimental Methods and results summarised in Table S4). Although Y6 and Y11 both have an A-DA'D-A structure, the central acceptor (A') groups differ between the two molecules with Y11 having benzotriazole (BTz) in the place of benzothiadiazole (BT) (Figure 1a). In Y6, the electron density around the central BT group is balanced by the peripheral A groups, resulting in a negligible dipole in the x-y plane. However, as BTz is less electron-withdrawing than BT [53,54], the dipole in Y11 is dominated by the peripheral A groups, resulting in a larger dipole in the x-y plane. Consequently, the intramolecular dipole is around a factor of four larger in Y11 (3.82 D) than in Y6 (Y6 = 0.97 D) (Figure S18). As a large intramolecular dipole can strengthen interactions between molecules and so provide a greater driving force for crystallisation [55-58], we ascribe enhanced ordering in PM6:Y11 to Y11's larger ground state dipole moment. It has been widely reported that more crystalline phases have increased triplet exciton diffusion lengths and mobilities, and so we propose that the enhanced long-range OOP ordering in PM6:Y11 is responsible for its increased rate of TTA [21-24,59].

*2.4. The Effect of TTA on EQE$_{EL}$ and $\Delta V_{nr}$*

As discussed in the introduction, TTA has the potential to improve an OSC's EQE$_{EL}$ (and thus reduce $\Delta V_{nr}$) by converting up to 50% of the optically dark triplet states formed via charge recombination into bright singlet states. Thus, to conclude this work, we estimate the amount by which TTA could reduce $\Delta V_{nr}$ to evaluate its potential impact on device performance. For this calculation, we assume that a fraction, ω, of the triplets formed by non-geminate recombination reform singlet states, with the rest decaying non-radiatively to the ground state. This gives rise to the cycle shown in Figure 5a, which indicates the possible decay routes of an injected polaron under the assumption of open circuit conditions (i.e., no net extraction of charge). To calculate the total probability of radiative decay, it is necessary to sum the contributions from the radiative decay of polarons which do not form triplet states (i.e., radiative decay via the spin-singlet CT state, $^1$CT) and the radiative decay of NFA singlet states reformed via TTA. As we do not know the precise mechanism by which TTA proceeds in these materials, we have performed this calculation for two different assumptions. In the first, we assume that all singlet excitons generated by TTA undergo radiative decay with no further polaron formation, giving

$$\chi = (1 - \alpha) + \alpha\omega \qquad (6)$$

This represents the best-case scenario and places an upper bound on the amount by which TTA could reduce $\Delta V_{nr}$. In the second, we assume that the singlet states generated by TTA may undergo charge separation and form polarons which will then loop around the cycle shown in Figure 5a until they decay (either radiatively or non-radiatively). In this case, χ is the sum of two infinite series and is given by

$$\chi = \frac{1 - \alpha + \alpha\omega(1 - \eta_{CT})}{1 - \alpha\omega\eta_{CT}} \qquad (7)$$

where $\eta_{CT}$ is the probability that a photogenerated singlet state dissociates to form polarons (see Figure S19 and S3.1). We consider that which of these descriptions is more accurate will depend upon the location in the NFA domains where TTA occurs (i.e., in the intermixed interfacial regions where singlet excitons formed by TTA are likely to re-dissociate into CT/charge separated states or in the pure NFA domains where the radiative decay of singlet excitons becomes more likely). The reduction in $\Delta V_{nr}$ which can be achieved by varying α and ω under each of these assumptions is shown in Figure 5b and

Figure 5c. These figures illustrate the range of potential $V_{OC}$ gains which are possible, which can be on the order of several tens of mV in the most favourable conditions. In both cases, the precise value of the reduction in $\Delta V_{nr}$ will depend not only upon the assumption made about the fate of the recycled singlet exciton states, but also upon the value ω, for which the upper limit is 0.5 as TTA always returns one triplet to the ground state (Figure 1b).

However, how closely ω can approach 0.5 is also determined by the kinetic competition between different possible triplet decay mechanisms (i.e., TTA, TCA and monomolecular triplet decay), which will depend upon both the rates of the individual processes and the incident photon flux as the flux determines the steady-state triplet and polaron populations. To understand this competition better, we illustrate in Figure 5d the fraction of triplet recombination which occurs via TTA as a function of TTA rate and triplet lifetime for a polaron population density of $4.5 \times 10^{16}$ cm$^{-3}$, approximately equal to that under 1-Sun conditions [60] (full details of the calculation are given in S3.2). We show here the case where the rate of TCA is low relative to the highest plotted values of the TTA rate constant, as our TAS and PLDMR results suggest that the right-hand side of this plot is representative of the parameters in the PM6:Y11 blend (plots with higher values of the TCA rate constant, indicative of the situation in PM6:Y6, are shown in Figure S20). Figure 5d indicates that a triplet lifetime greater than ~ 1 microsecond is necessary for TTA to be a significant decay pathway under 1-Sun conditions, whereas our combined TTA-monomolecular fit to the triplet decay in the neat Y11 film data suggests that the Y11 monomolecular triplet lifetime is 63 ± 7 ns (see Table S1 and S1.2). We note that this value is extremely low compared to previous reports of monomolecular triplet lifetimes in other NFA films, which can be up to ~100 microseconds for some ITIC derivatives [61]. Thus, whilst we find a greatly reduced non-radiative voltage loss in the PM6:Y11 blend of 220 mV compared to 260 mV in PM6:Y6 (Figure S21), our analysis suggests that TTA is unlikely to be the dominant triplet decay mechanism in PM6:Y11 OSCs under 1-Sun illumination. This implies that other factors, such as a smaller $S_1$-CT offset, or increased $S_1$-CT hybridisation, may be responsible for the reduced non-radiative losses in PM6:Y11 [62 - 64]. However, for material systems where the rate of TTA significantly exceeds that of TCA and the monomolecular triplet lifetime is on the order of 1 microsecond or greater, TTA could make an important contribution to reducing non-radiative voltage losses associated with the recombination of free charge carriers to triplet excitons.

*3. Conclusion*

In this work, we have combined the results of PLDMR and kinetic modelling of fluence-dependent TAS data to conclude that TTA is the dominant triplet decay mechanism in PM6:Y11 at high fluences. This contrasts with PM6:Y6, where triplet decay is dominated by TCA. We attribute this discrepancy between the blends to the difference in the ordering of the NFA domains, as revealed by our PLDMR and GIWAXS measurements. These showed that the NFA domains in PM6:Y11 have a greater degree of long-range OOP ordering than those of PM6:Y6, which is driven by the larger ground state molecular dipole moment of Y11. The enhanced crystallinity of PM6:Y11 when compared to PM6:Y6 suggests that NFA triplet excitons can exhibit higher mobilities in the former blend, leading to a higher rate of TTA.

We have also investigated the potential impact of TTA on the value of $\Delta V_{nr}$ and concluded that, for TTA to significantly improve $V_{OC}$ under 1-Sun conditions, the triplet lifetime must be greater than 1 microsecond as otherwise the triplet population cannot reach sufficiently high densities for TTA, a second order decay process, to compete with monomolecular triplet decay. As our estimates of the triplet lifetimes in neat films of Y6 and Y11 are in the range of 10s-100s of nanoseconds, we conclude that TTA is unlikely to significantly affect device performance in the PM6:Y6 and PM6:Y11 blends studied herein. However, if the triplet lifetime of the NFA could be extended while maintaining a rate of TTA which is significantly higher than that of TCA, we have demonstrated that the non-radiative voltage losses could be reduced by tens of mV.

Consequently, we can provide important design rules for encouraging TTA in NFA OSCs. First, it is important to maximise the intrinsic TTA rate of the NFA component, which we find can be achieved by increasing the molecular ordering of the NFA domains. Thus, we propose that modification of the NFA molecular structure and side chains to increase the ground state dipole moment and crystallinity [65], as well as the use of film processing steps like thermal annealing [50] and solvent additives [66] to modulate the NFA packing, could be important tools to influence TTA. Second, it is critical to ensure that the triplet lifetime is long enough for a significant population of triplet excitons to build up under 1-Sun excitation. However, this requirement must be carefully balanced against the potential for triplet excitons to interact with molecular oxygen and initiate material degradation [67]. Our results suggest that engineering an NFA triplet exciton lifetime on the order of 1 microsecond could provide a good balance between encouraging TTA whilst minimising the risk of unwanted side reactions. Thus, we believe that with further study and refinement, TTA could be an important tool in mitigating against non-radiative triplet losses in OSCs.

**Tables**

**Table 1.** Fit parameters from the global analysis of the PM6:Y6 and PM6:Y11 femtosecond TAS data. See S1.1 for a detailed discussion of the fitting procedure and assumptions.

| **Material** | **A** | **B [s$^{-1}$]** [a] | **G [s$^{-1}$]** [a] |
|---|---|---|---|
| PM6:Y6 | 0.299 ± 0.005 | (6.3 ± 0.6) × 10$^{10}$ | (7 ± 2) × 10$^{10}$ |
| PM6:Y11 | 0.59 ± 0.03 | - | (2.3 ± 0.2) × 10$^{11}$ |

[a] The units of $B$ and $G$ are s$^{-1}$, not cm$^3$ s$^{-1}$, due to the normalisation procedure used to carry out the 'cross-section free' fits (see equation S3).

# Figures

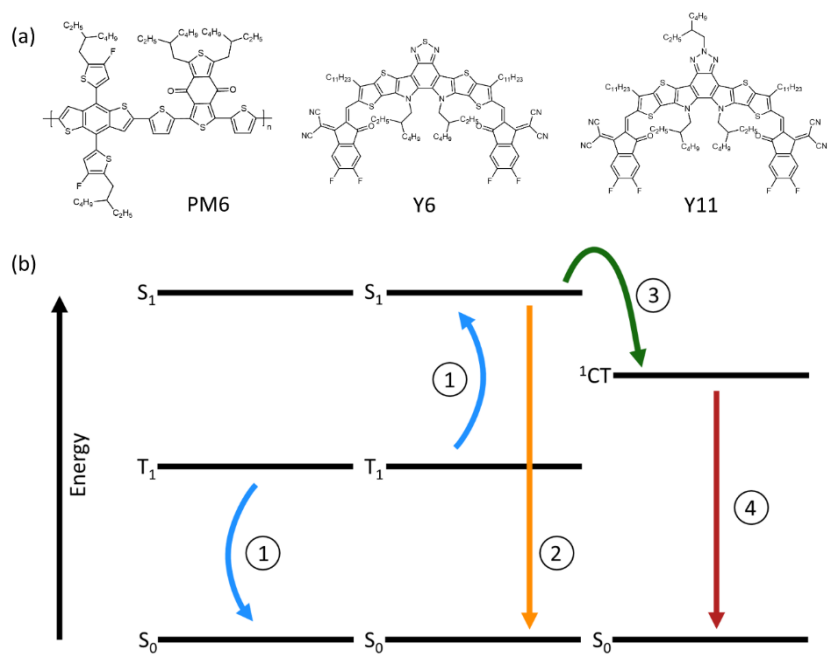

**Figure 1:** (a) The chemical structures of the materials discussed in this paper. PM6 acts as the donor in both device structures and the acceptor is either Y6 or Y11. (b) A schematic diagram of triplet-triplet annihilation (TTA) and the mechanism by which it could increase $EQE_{EL}$. The highlighted processes are: 1) TTA in which two Y11 triplet excitons ($T_1$) on neighbouring molecules interact, resulting in one molecule returning to the ground state ($S_0$) and the other being excited to the singlet state ($S_1$). 2) The $S_1$ state decays radiatively to the ground state, emitting a photon. 3) The $S_1$ state forms a $^1CT$ state at an interface between PM6 and Y11. 4) The $^1CT$ state decays radiatively to the ground state, emitting a photon. TTA is hypothesised to increase $EQE_{EL}$ as it increases the fraction of excited states which can decay radiatively (processes 2 and 4) by forming a bright $S_1$ state from two, non-radiative $T_1$ states (process 1).

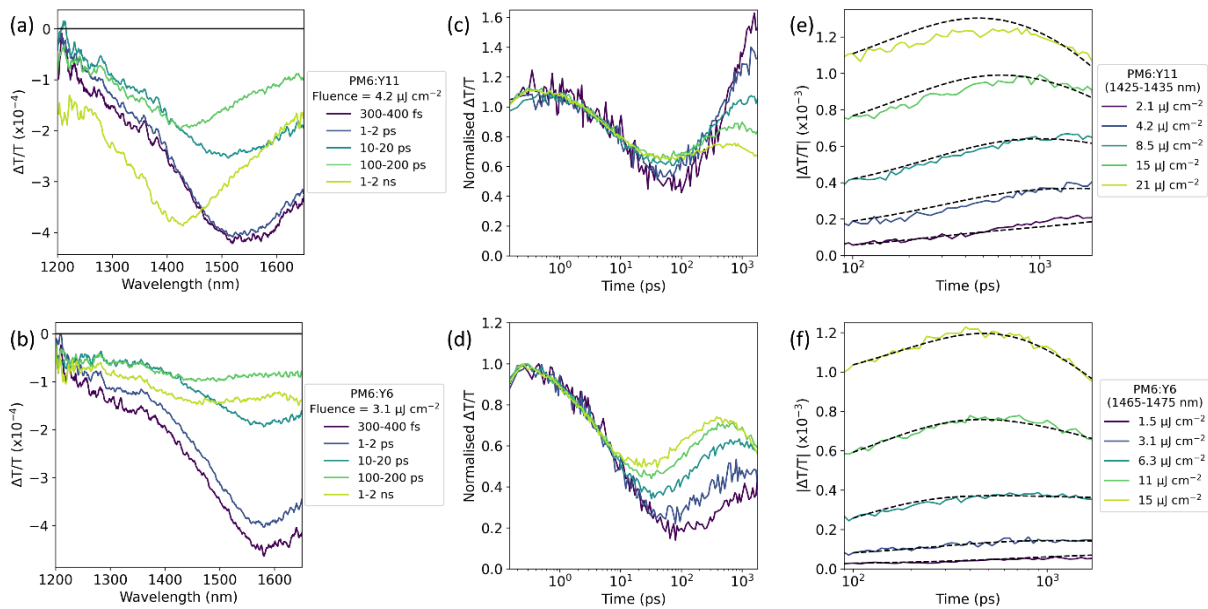

**Figure 2:** (a) Transient Absorption (TA) spectra for PM6:Y11 in the IR spectral region. The pump wavelength was 800 nm, preferentially exciting the Y11. Figure 2b shows the same measurement performed on PM6:Y6. We identify two signals in each spectra: an inter-CT state PIA in the region 1500-1600 nm and a triplet exciton PIA in the region 1400-1500 nm. The triplet exciton PIA appears weaker relative to the inter-CT state PIA in PM6:Y6 than in PM6:Y11 in part due to the higher formation yield of the inter-CT state from singlet excitons in the former (see Figure S4) The full TA spectra for both blends in the visible and NIR/IR regions are given in Figure S3, where the most prominent spectral features are identified. (c) Fluence series of the Y11 triplet kinetic taken over the wavelength range 1425-1435 nm. The fluence dependence of the signals' rise from ~100 ps onwards indicates that it is caused by a bi-molecular (or higher order) process, suggesting that triplet formation occurs via a non-geminate pathway. (d) Fluence series of the Y6 triplet kinetic taken over the wavelength range 1465-1475 nm. Note how, at higher fluences, the triplet maximum increases relative to the point of normalisation, in contrast to the behaviour of the Y11 triplet. (e) Results of the global fit for the Y11 triplet population. (f) Results of the global fit for the Y6 triplet population.

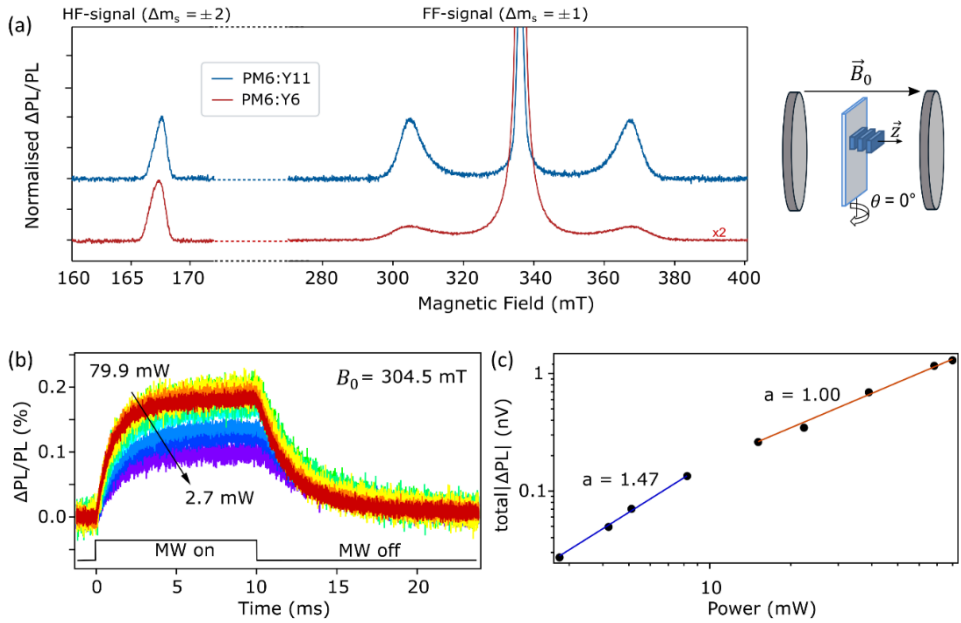

**Figure 3:** (a) Photoluminescence detected magnetic resonance (cwPLDMR) of PM6:Y11 (blue) and PM6:Y6 (red) recorded at T = 10 K. The spectral width of the full-field (FF) signal and the position of the half-field (HF) signal allow us to assign the broad spectral feature to triplet excitons on the NFA. At $\theta = 0°$ (where $\theta$ is the angle between the molecular z-axis and the external magnetic field, see inset) the spectra reveal a preferential orientation due to intermolecular face-on stacking and face-on stacking on the substrate. The wings of PM6:Y11 are steeper than PM6:Y6 (with ordering factors of $\lambda_\theta = 9.0$ and $\lambda_\theta = 5.5$ respectively), indicating that PM6:Y11 has higher crystallinity in the OOP direction. Figures 3b-c show transient PLDMR (trPLDMR) of PM6:Y11 for different laser excitation powers at the position of the triplet feature (B = 304.5 mT, see Figure 3a). (b) The PLDMR transients increase (decrease) in intensity upon switching the microwave (MW) field on (off). Signal saturation is reached within several ms. The PL enhancement ($\Delta$PL/PL) of the transients' triplet feature increases upon increasing the laser excitation power, until it reaches a maximum at laser excitation powers above 15 mW. (c) The laser excitation dependence of absolute PLDMR signal ($\Delta$PL), used to determine the origin of triplet-sensitive PL enhancement. The data were fitted to the power law $\Delta PL \sim P_{exc}^{a}$ and two distinct regimes were identified. In the low-power regime (below ~10 mW) $a = 1.47$. However, upon increasing the power, the value of a decreased to $a = 1.00$. This excitation power dependence is typical for TTA upconversion systems, which display an annihilation-limited regime at higher powers. For comparison, the laser excitation dependence of $\Delta$PL/PL is shown in Figure S15, which demonstrates its independence from the laser excitation intensity above ~ 10 mW.

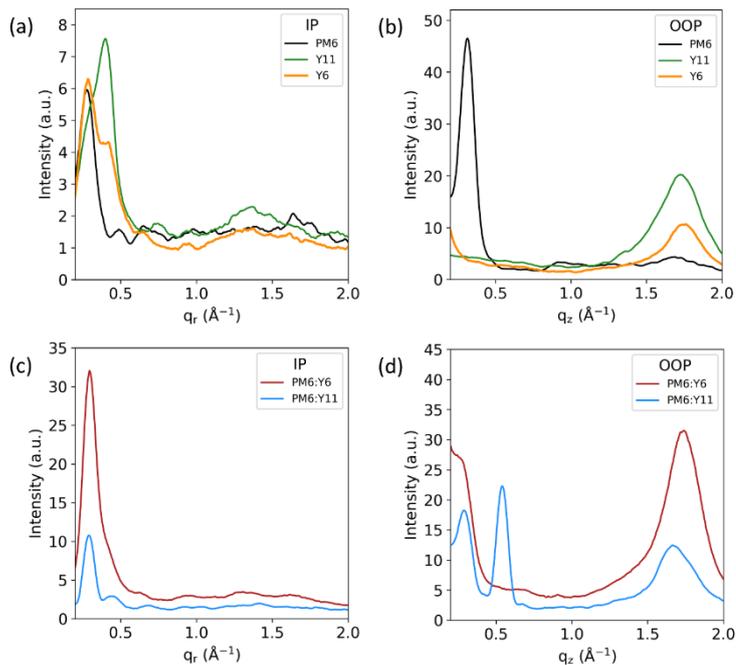

**Figure 4:** GIWAXS measurements performed on neat PM6, Y6 and Y11 films and the blended PM6:Y11 and PM6:Y6 films. Figures (a) and (b) show the line cuts for the neat films in the IP and OOP directions respectively and Figures (c) and (d) show the line cuts for the blend films in the IP and OOP directions respectively. In the PM6:Y11 bend, an additional OOP (100) peak is observed at q~0.6 Å$^{-1}$, indicating an enhanced crystallinity of the Y11 domains in the blended film. Full 2D GIWAXS images are given in Figure S17 and the d-spacings of the peak locations are given in Table S3.

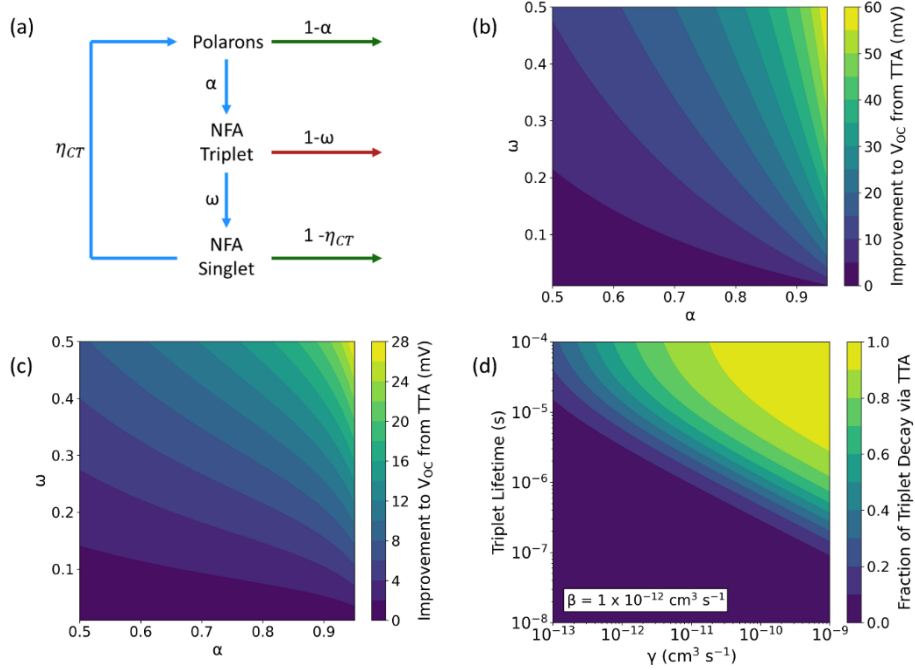

**Figure 5:** Calculations of the fraction of triplet decay that occurs via TTA and possible improvement in $V_{OC}$ due to TTA. For full details of the calculations and assumptions, see the main text and S3. (a) A cycle representing the possible decay channels for an injected polaron at an injected current density equivalent to open circuit conditions. $\eta_{CT}$ is the probability that a singlet exciton dissociates to form polarons and has assumed to take a value of 0.93, as calculated from the PM6:Y6 TAS data (see S3.1) and consistent with the high IQE values reported for PM6:Y6 devices [20,21]. Green arrows indicate processes that could lead to radiative decay and red arrows those which could not. (b) A plot of the improvement to $V_{OC}$ due to the presence of TTA as a function of α and ω, assuming that each singlet exciton produced by TTA immediately undergoes radiative decay (i.e., $\eta_{CT} = 0$). (c) A plot of the improvement to $V_{OC}$ due to the presence of TTA as a function of α and ω, assuming that singlet excitons produced by TTA undergo charge separation with probability $\eta_{CT}$, creating polarons which can then go around the cycle shown in (a) indefinitely prior to decay. In both (b) and (c), the value of α is capped at 0.95 because the possible improvement from TTA becomes infinite if we assume that all non-geminate recombination forms triplet states. (d) The fraction of the triplet states which decay via TTA under polaron densities equivalent to 1-Sun conditions and a TCA rate constant, β, of $1 \times 10^{-12}$ cm$^3$s$^{-1}$. The value of β was chosen to be indicative of the situation in PM6:Y11.

**Experimental Methods**

*Film and Device Fabrication:* OSCs were fabricated in the configuration of the traditional sandwich structure with an indium tin oxide (ITO) glass positive electrode and PDINO/Al negative electrode. ITO-coated glass substrates were rinsed with deionized water, acetone and isopropyl alcohol by ultrasonication, sequentially and then dried with nitrogen. A thin layer of PEDOT:PSS (poly(3,4-ethylenedioxythiophene): poly(styrene sulfonate)) was prepared by spin-coating the PEDOT:PSS solution filtered through a 0.45 mm poly(tetrafluoroethylene) (PTFE) filter at 3,000 rpm for 40 s on the ITO substrate. Subsequently, PEDOT:PSS film was baked at 150 °C for 15 min in the air, and the thickness of the PEDOT:PSS layer was about 40 nm. The PM6:Y6 (D:A=1:1.2, 16 mg mL$^{-1}$ in total) and PM6:Y11 (D:A=1:1.5, 16 mg mL$^{-1}$ in total) were dissolved in chloroform with the solvent additive of 1-chloronaphtalene (CN) (0.5 %, v/v) and spin-cast at 3,000 rpm for 30 s onto the PEDOT:PSS layer. A bilayer cathode consisting of PDINO (~15 nm) capped with Al (~150 nm) was thermally evaporated under a shadow mask with a base pressure of ca. 10$^5$ Pa. Finally, top electrodes were deposited in a vacuum onto the active layer. The active area of the device was 5 mm$^2$.

Thin film samples for optical measurements were prepared with the same solutions and treatments for device fabrication on quartz substrates.

All the devices and films were fabricated in a nitrogen-filled glove box.

*Device Characterisation*: All device characterisation was carried out in a nitrogen-filled glovebox.

J-V characterisation was carried out under AM 1.5G irradiation with the intensity of 100 mW cm$^{-2}$ (Oriel 67005, 500 W), calibrated by a standard silicon cell. J-V curves were recorded with a Keithley 236 digital source meter. A xenon lamp with AM 1.5 filter was used as the white light source and the optical power was 100 mW cm$^{-2}$.

The EQE$_{PV}$ measurements were performed using a Stanford Systems model SR830 DSP lock-in amplifier coupled with a WDG3 monochromator and 500 W xenon lamp. A calibrated silicon detector was used to determine the absolute photosensitivity at different wavelengths.

EQE$_{EL}$ values were obtained from an in-house-built system including a Hamamatsu silicon photodiode 1010B, a Keithley 2400 SourceMeter to provide voltage and record injected current, and a Keithley 485 Picoammeter to measure the emitted light intensity. The system was calibrated within the detecting range of silicon.

*FTPS-EQE*: FTPS-EQE was measured using Vertex 70 from Bruker Optics, equipped with a quartz tungsten halogen lamp, quartz beam splitter and external detector option. A low-noise current amplifier (SR570) was used to amplify the photocurrent produced on illumination of the photovoltaic devices with light modulated by the Fourier transform infrared spectroscope (FTIR). The output voltage of the current amplifier was fed back into the external detector port of the FTIR, to be able to use the FTIR's software to collect the photocurrent spectrum.

*Photoluminescence quantum yield measurements*: Photoluminescence quantum yield was performed in an N-M01 integrating sphere from Edinburgh Instruments. Spectra were recorded by a Newton EM-CCD Si array detector cooled at -45 °C with a Shamrock SR-303i spectrograph from Andor Tech. Indirect excitation emissions were subtracted for the absolute quantum yield calculation.

*Transient Absorption Spectroscopy*: TAS was performed on either one of two experimental setups. The femtosecond TAS in the IR region (900 – 1650 nm) was performed on a setup powered using a commercially available Ti:sapphire amplifier (Spectra Physics Solstice Ace). The amplifier operates at 1 kHz and generates 100 fs pulses centred at 800 nm with an output of 7 W. A portion of the laser fundamental was used for sample excitation at 800 nm. For the nanosecond TAS measurements, the probe was generated by a LEUKOS Disco 1 UV low timing jitter supercontinuum laser (STM-1-UV), which was then electronically delayed relative to the femtosecond 800 nm excitation by an electronic delay generator (Stanford Research Systems DG645). The probe pulses are collected with an InGaAs dual-line array detector (Hamamatsu G11608-512DA), driven and read out by a custom-built board from Stresing Entwicklungsbüro. The probe beam was split into two identical beams by a 50/50 beamsplitter. This allowed for the use of a second reference beam which also passes through the sample but does not interact with the pump. The role of the reference was to correct for any shot-to-shot fluctuations in the probe that would otherwise greatly increase the structured noise in our experiments.

For the 500 – 950 nm continuous probe region TAS, a Yb amplifier (PHAROS, Light Conversion), operating at 38 kHz and generating 200 fs pulses centred at 1030 nm with an output of 14.5 W was used. The ~200 fs pump pulse was provided by an optical parametric amplifier (Light Conversion ORPHEUS). The probe is provided by a white light supercontinuum generated in a YAG crystal from a small amount of the 1030 nm fundamental. After passing through the sample, the probe is imaged using a Si photodiode array (Stresing S11490).

*Photoluminescence Detected Magnetic Resonance*: PLDMR experiments were carried out with a modified X-band spectrometer (Bruker E300) equipped with a continuous-flow helium cryostat (Oxford ESR 900) and a microwave cavity (Bruker ER4104OR, ~9.43 GHz) with optical access. Optical irradiation was performed with a 532 nm continuous wave laser (Cobolt Samba CW 532 nm DPSSL) from one side opening of the cavity. PL was detected with a silicon photodiode (Hamamatsu S2281) on the opposite opening, using a 561 nm longpass filter to reject the excitation light. The PL signal was amplified by a current/voltage amplifier (Femto DHPCA-100). For cwPLDMR, PL was recorded by a lock-in detector (Ametek SR 7230), referenced by on-off modulating the microwaves with a modulation frequency of 547 Hz. The microwaves were generated with a microwave signal generator (Anritsu MG3694C), amplified to 3 W (microsemi) and guided into the cavity. For trPLDMR, PL was recorded by a digitizer card (GaGe Razor Express 1642 CompuScope), whereby a pulse blaster card (PulseBlasterESR-PRO) triggered the digitizer card and the microwave generator to produce microwave pulses for a set length. The microwave pulses were amplified to 5 W by a traveling wave tube amplifier (TWTA, Varian VZX 6981 K1ACDK) and guided into the cavity.

*GIWAXS*: GIWAXS measurements were carried out with a Xeuss 2.0 SAXS/WAXS laboratory beamline using a Cu X-ray source (8.05 keV, 1.54 Å) and a Pilatus3R 300K detector. The incidence angle is 0.2. All measurements were conducted under a vacuum environment to reduce air scattering.

*Density Function Theory Simulations:* Single-molecule gas-phase DFT simulations were performed using Gaussian16 software on the Imperial College High Performance Computing service, with GaussView 6 used for result visualization. DFT was applied at the B3LYP level of theory with the 6-311G(d,p) basis set. The calculations are carried out on molecules with full side chains.

**Supporting Information**

Supporting Information is available on arXiv.

## Conflict of Interest Statement

The authors declare that no conflict of interest exists.

## Data Availability Statement

The data that support the plots within this paper is available at the University of Cambridge Repository (to be completed in proofs).

## Author Contributions

A.J.G. and J.Y. conceived the work. A.J.G. performed the TAS measurements. L.J.F.H. analysed the TAS data and developed the fitting code and the TTA recombination cycle model. J.G. carried out the PLDMR measurements. J.Y, W.L. and H. Z. synthesised the NFAs, fabricated and tested the OSC devices, and performed the PLQY measurements. T.L. performed the GIWAXS and X.J. assisted with the data interpretation. J.L. and Y.-C.C. performed the dipole moment calculations. H.Z. made the samples for TAS. D.J.C.S. and D.M.L.U. fabricated the samples for the PLDMR experiments. J.-S.K., Y.Z., X.L, F.G., A.S., V.D., J.Y., and A.J.G. supervised their group members involved in the project. L.J.F.H., J.G., and A.J.G. wrote the manuscript with input from all authors.

## Acknowledgements

A.J.G. thanks the Leverhulme Trust for an Early Career Fellowship (ECF-2022-445). L.J.F.H. thanks the UK Engineering and Physical Sciences Research Council (EPSRC) Application Targeted and Integrated Photovoltaics (ATIP) project (EP/T028513/1) for support. J. Y. acknowledge the National Natural Science Foundation of China (No. 22005347). Y. Z. acknowledge the National Natural Science Foundation of China (No. 521253). J.G., A.S., and V.D. acknowledge support by the Deutsche Forschungsgemeinschaft (DFG, German Research Foundation) within the Research Training School "Molecular biradicals: Structure, properties and reactivity" (GRK2112) and the Bavarian Ministry of the Environment and Consumer Protection, the Bavarian Network "Solar Technologies Go Hybrid". X.J. acknowledges the China Scholarship Council (CSC) for funding. T.L. and X.L. acknowledge the Research Grant Council of Hong Kong (No. 14303519). We thank Professor Sir Richard H. Friend for his insights and many useful discussions.

# Supplementary Information for, 'Understanding the Role of Triplet-triplet Annihilation in Non-fullerene Acceptor Organic Solar Cells'

Lucy J. F. Hart[1,2], Jeannine Grüne[1,3], Wei Liu[4], Tsz-ki Lau[5], Joel Luke[6], Yi-Chun Chin[6], Xinyu Jiang[7], Huotian Zhang[8], Daniel J.C. Sowood[1], Darcy M. L. Unson[1], Ji-Seon Kim[6], Xinhui Lu[5], Yingping Zou[4], Feng Gao[8], Andreas Sperlich[3], Vladimir Dyakonov[3], Jun Yuan[4*] and Alexander J. Gillett[1*].

[1]Cavendish Laboratory, University of Cambridge, JJ Thomson Avenue, Cambridge, U.K.
[2]Department of Chemistry and Centre for Processable Electronics, Imperial College London, 82 Wood Lane, London, U.K.
[3]Experimental Physics 6, Julius Maximilian University of Würzburg, Am Hubland, Würzburg 97074, Würzburg, Germany
[4]College of Chemistry and Chemical Engineering, Central South University, Changsha 410083, P.R. China.
[5]Department of Physics, The Chinese University of Hong Kong, Shatin, 999077 Hong Kong.
[6]Department of Physics and Centre for Processable Electronics, Imperial College London, South Kensington, U.K.
[7]Chair for Functional Materials, Department of Physics, TUM School of Natural Sciences, Technical University of Munich, James-Franck-Str. 1, 85748 Garching, Germany.
[8]Department of Physics, Chemistry and Biology (IFM), Linköping University, Linköping, Sweden.

*Corresponding authors: Alexander J. Gillett: E-mail: ajg216@cam.ac.uk; Jun Yuan: E-mail: junyuan@csu.edu.cn.

**Supplementary Tables**

| Material | τ (ns) | G (s$^{-1}$) |
|---|---|---|
| Y6 | 600 ± 400 | (1.20 ± 0.04) × 10$^{11}$ |
| Y11 | 63 ± 7 | (1.46 ± 0.04) × 10$^{11}$ |

**Table S1:** Fit parameters obtained from the global analysis of the neat Y6 and neat Y11 nanosecond TAS data (see S1.2 for details of the fitting procedure).

| | | **Triplet** | | | | **CT** | |
|---|---|---|---|---|---|---|---|
| Material | Orient. | $D, E$ (MHz) | $\lambda_\theta, \lambda_\varphi$ | Lw (mT) | weight | Lw (mT) | weight |
| PM6:Y11 | 0° | 930, 140* | 9.0, 0 | 6.0, 0 | 0.57 | 0.9, 2.0 | 0.43 |
| PM6:Y11 | 45° | 930, 140 | 0.0, 0 | 6.0, 0 | 0.37 | 0.9, 1.7 | 0.63 |
| PM6:Y6 | 0° | 990, 140* | 5.5, 0 | 8.0, 0 | 0.12 | 1.1, 2.5 | 0.88 |
| PM6:Y6 | 45° | 990, 140 | 0.0, 0 | 8.0, 0 | 0.10 | 1.1, 2.2 | 0.90 |
| Y11 | 0° | 950, 140* | 11.0, 0 | 3.7, 0 | 0.74 | 0, 3.0 / 0, 1.3 | 0.26 / -0.18 |
| Y6 | 0° | 950, 150* / 950, 150 | 11.0, 0 / 0.0, 0 | 4.0, 0 / 2.0, 0 | 0.62 / 0.11 | 1, 2.1 / 0, 1.3 | 0.27 / -0.06 |
| PM6 | 0° | 1500, 70* / 1500, 70 | 5.5, 0 / 0, 2.0 | 9.0, 0 / 5.0, 0 | 0.30 / 0.51 | 0, 1.7 | 0.19 |

**Table S2:** Parameters for PLDMR spectral simulations using the MATLAB toolbox EasySpin. *: $E$ value cannot be determined due to high ordering. Linewidth (Lw) given in Gaussian, Lorentzian.

| Material | IP (100) - Å | IP (010) - Å | OOP (100) - Å | OOP (010) - Å |
|---|---|---|---|---|
| PM6 | 21.00 | - | 21.00 | - |
| Y11 | 15.30 | - | - | 3.63 |
| PM6:Y11 | 21.66 | - | 21.66 / 11.56 | 3.74 |

**Table S3:** Calculated d-spacings of the Bragg peaks which are present in GIWAXS data shown in Figure 5. The d-spacings of the peaks are estimated using d = 2πq$^{-1}$, where q refers to the Bragg peak position.

| Material | Dihedral Angle | Dipole (D) | $D_x$ (D) | $D_y$ (D) | $D_z$ (D) |
|---|---|---|---|---|---|
| Y6 | 8.5° | 0.97 | 0.01 | 0.02 | 0.97 |
| Y11 | 7.1° | 3.82 | 0.32 | 3.78 | 0.42 |

**Table S4:** Equilibrium geometry dihedral angles and dipoles extracted from density functional theory simulations of Y6 and Y11. The directions of $D_x$, $D_y$ and $D_z$ are defined in Figure S18.

# Supplementary Figures

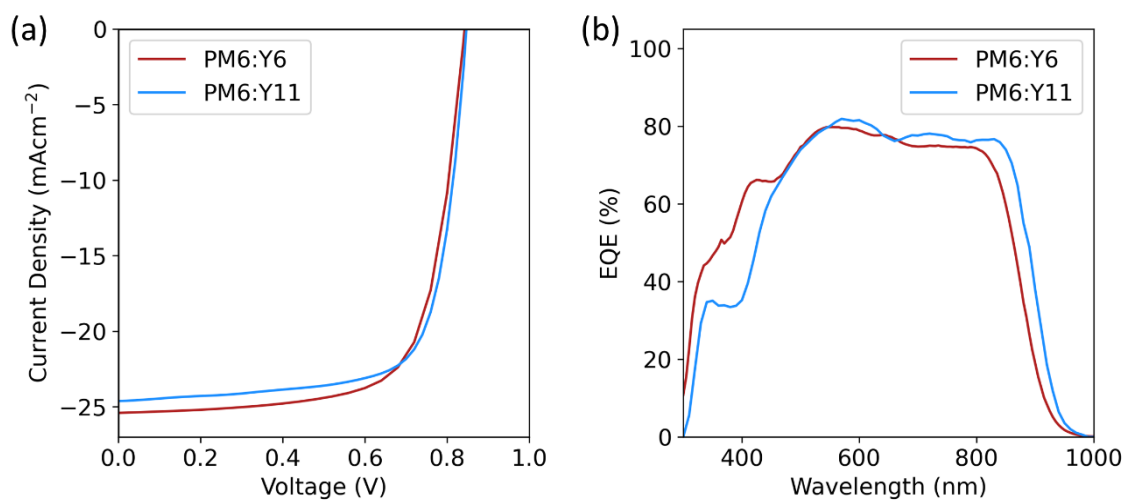

**Figure S1:** The (a) JV curves and (b) EQE$_{PV}$ of PM6:Y11 and PM6:Y6 organic solar cells whose active layers were prepared under the same conditions as those used to make the blend films measured in this work. For PM6:Y6, the device parameters are: V$_{OC}$ = 0.84 V, J$_{SC}$ = 25.4 mAcm$^{-2}$, FF = 0.71 and PCE = 15.2% and for PM6:Y11 they are: V$_{OC}$ = 0.85 V, J$_{SC}$ = 24.5 mAcm$^{-2}$, FF = 0.73 and PCE = 15.1%.

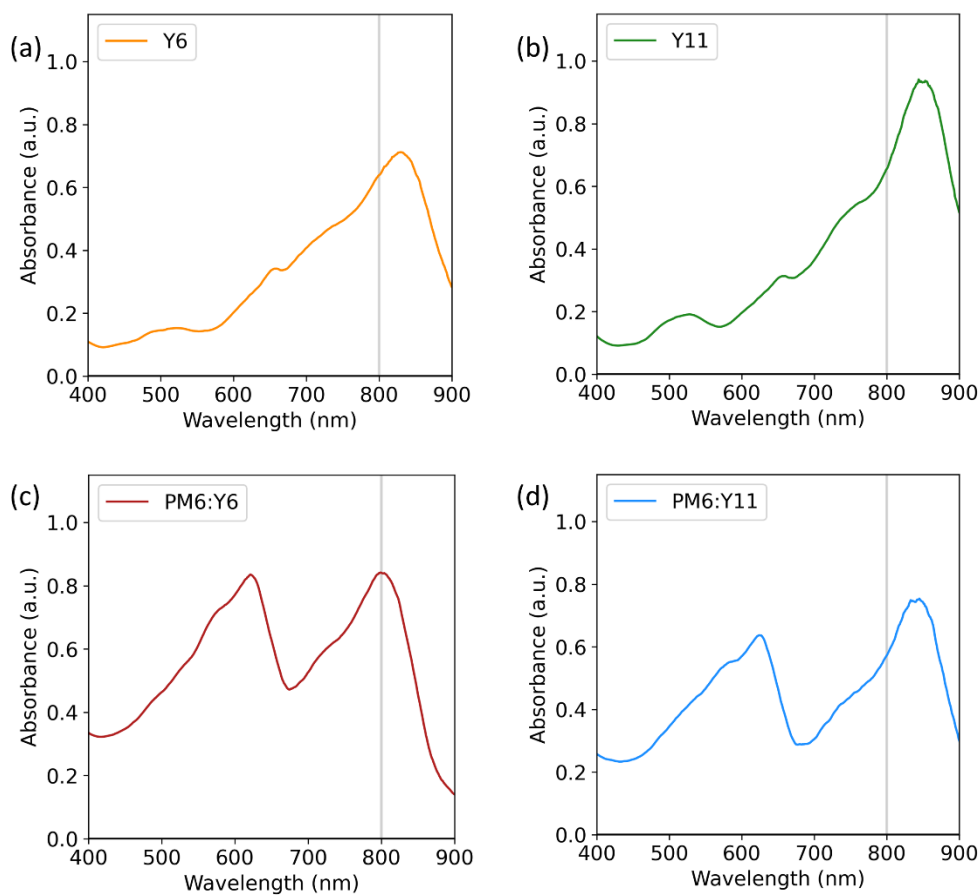

**Figure S2:** The absorption spectra of the (a) neat Y6, (b) neat Y11, (c) PM6:Y6 and (d) PM6:Y11 films used in this work. The grey shaded region indicates the absorbance at 800 nm, the wavelength used by the TAS pump laser.

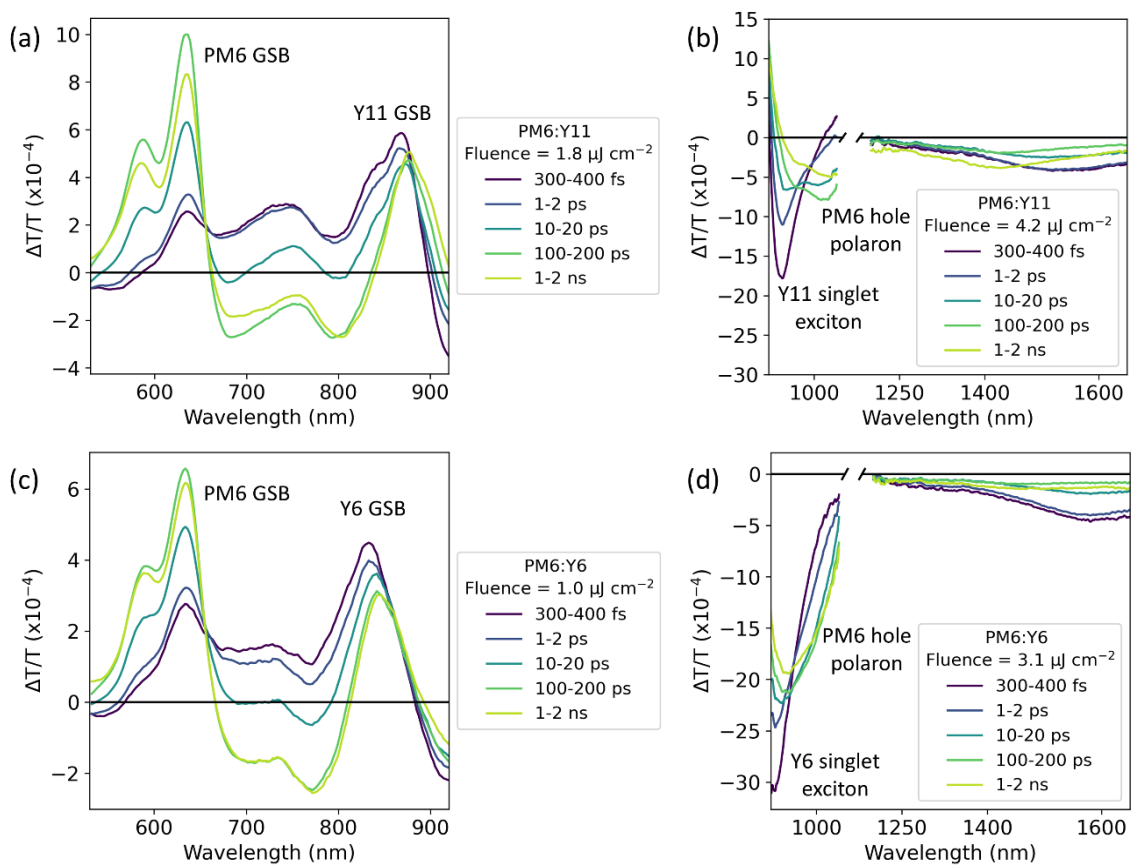

**Figure S3:** The visible, NIR and IR femtosecond TAS spectra of both PM6:Y11 (Figures a-b) and PM6:Y6 (Figures c-d) following excitation at 800 nm. The left column shows the visible region, while the right shows the NIR and IR regions. The IR region for both blends is shown in greater detail in Figure 2.

We identify the ground state bleach (GSB) of the NFAs and PM6 by comparison to the absorption spectra of the neat materials (the NFA absorption spectra are shown in Figure S2 and see e.g., Gillett et al. for that of PM6 [1]). The spectra of the neat NFA films (Figure S4) have two distinct photoinduced absorption (PIA) peaks at early times: one in the 900-950 nm region and another in the 1500-1600 nm region. Both of these PIAs are also visible in the spectra of the blended films, shown here, allowing us to assign them to excited states on the NFA. Following recent reports in the literature, we identify the 900-950 nm region PIA as the NFA singlet exciton and the 1500-1600 nm PIA as a delocalised, inter-CT state on the NFA [2]. Finally, we assign the broad PIA which emerges around 100 ps in the 900-1000 nm region to the PM6 hole polaron by reference to the kinetics shown in Figure S5.

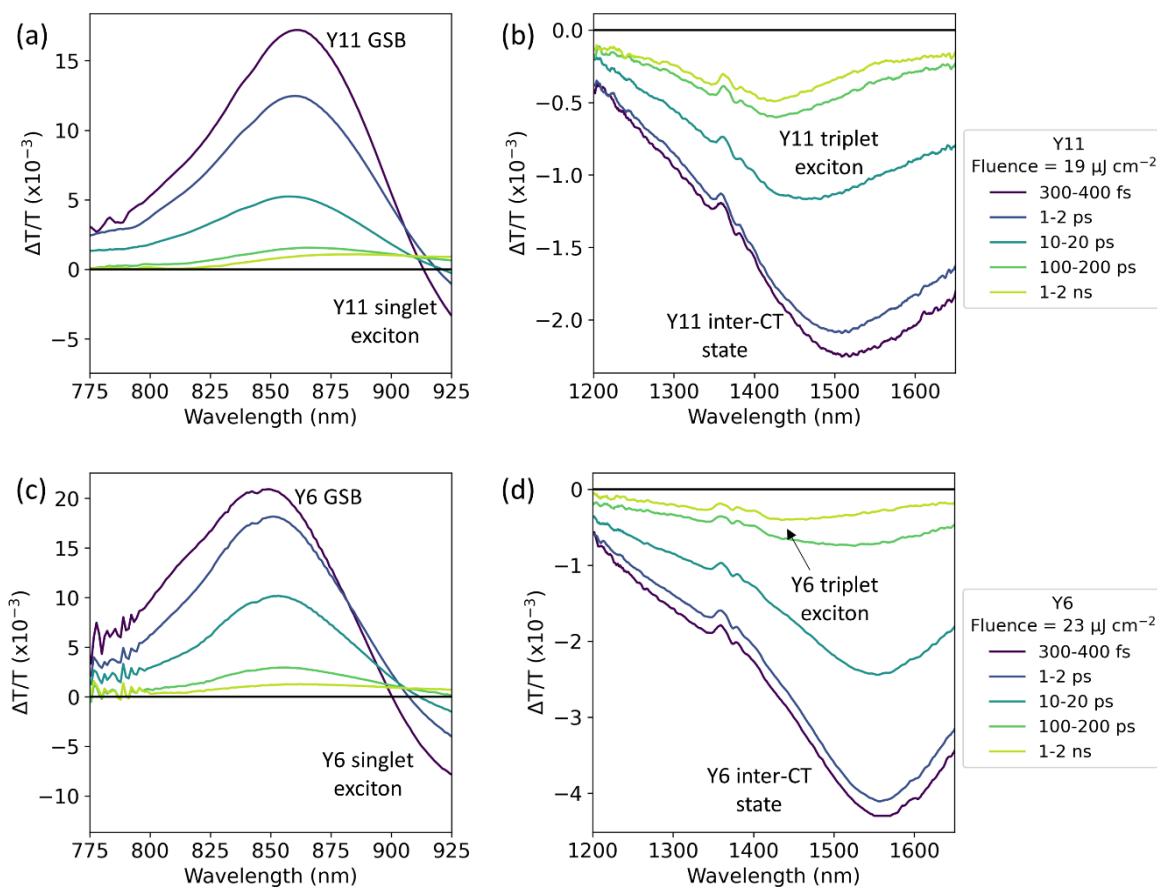

**Figure S4:** The NIR and IR femtosecond TAS spectra of both neat Y11 (Figures a-b) and neat Y6 (Figures c-d) following excitation at 800 nm. Figures a and c show the NIR region, while Figures b and d show the IR region. The signal strength is lower for the Y6 film despite the higher excitation fluence due to its lower absorbance (see Figure S2). A high fluence is shown here for so that the triplet exciton feature is clearly defined as, in the neat films, triplet excitons can only be formed by inter-system crossing and so have a relatively low yield.

By comparing the ratio of the GSB and the inter-CT state at 0.3 - 0.4 ps (7.6 in neat Y11 versus 4.9 in neat Y6) we find that this value is 1.6 times larger in neat Y11. If we assume that the ratio of the GSB to the inter-CT PIA absorption cross sections are the same in neat Y6 and neat Y11, this indicates that Y6 has a higher yield of singlet to inter-CT states than Y11.

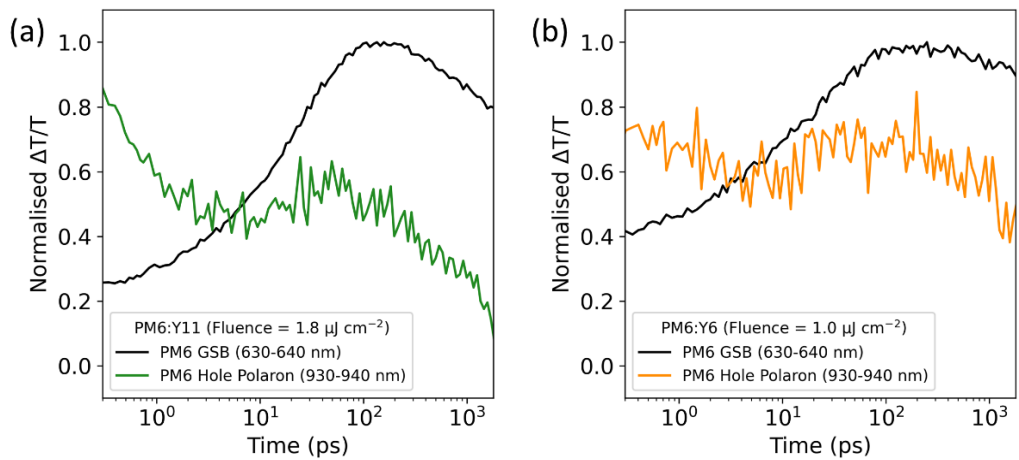

**Figure S5:** The femtosecond TAS kinetics of (a) PM6:Y11 and (b) PM6:Y6, illustrating both the PM6 GSB and the PM6 hole polaron PIA (see Figure S3a and S3c for the relevant spectra). As both blends were pumped at 800 nm, the NFA component has been selectively excited and thus the dominant charge transfer process is hole transfer from the NFA to the PM6. We ascribe the slow rise of the PM6 GSB to this hole transfer process. After around 100 ps, the PM6 GSB reaches a maximum in both blends, indicating that hole transfer is complete, and no new excited species are being created on the PM6. We now consider the 930-940 nm PIA, which is convoluted with the NFA singlet exciton PIA at times < 100 ps, when charge transfer is not yet complete (see Figure S3). However, beyond this time, it is clear that the decay of the PIA starts to mirror the decay of the PM6 GSB. As any PM6 singlet excitons generated by the pump laser will have undergone rapid charge transfer to the NFA and assuming a negligible PM6 triplet exciton population [1], the only excited species left on the PM6 after 100 ps are hole polarons. As such, we assign the 930-940 nm PIA to this species. Femtosecond TAS measurements on PM6:PCBM blend films agree with this assignment [1].

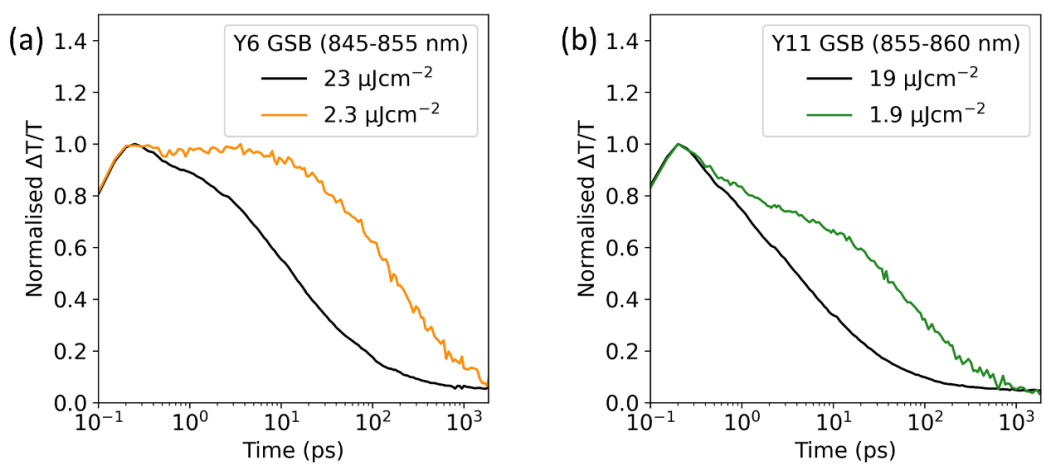

**Figure S6:** The GSB of (a) neat Y6 and (b) neat Y11 films following excitation at 800 nm. Both signals do not return to a $\Delta T/T$ value of zero at late times, indicating a long-lived excited state population. By examining the IR region of the spectrum (Figure S4), the remaining species can be identified as triplet exciton states, which are generated by inter-system crossing. Additionally, the fluence dependence of the decay at early times (<1 ps) indicates the presence of significant singlet-singlet (or inter-CT-inter-CT) annihilation processes, even at a relatively low excitation fluence.

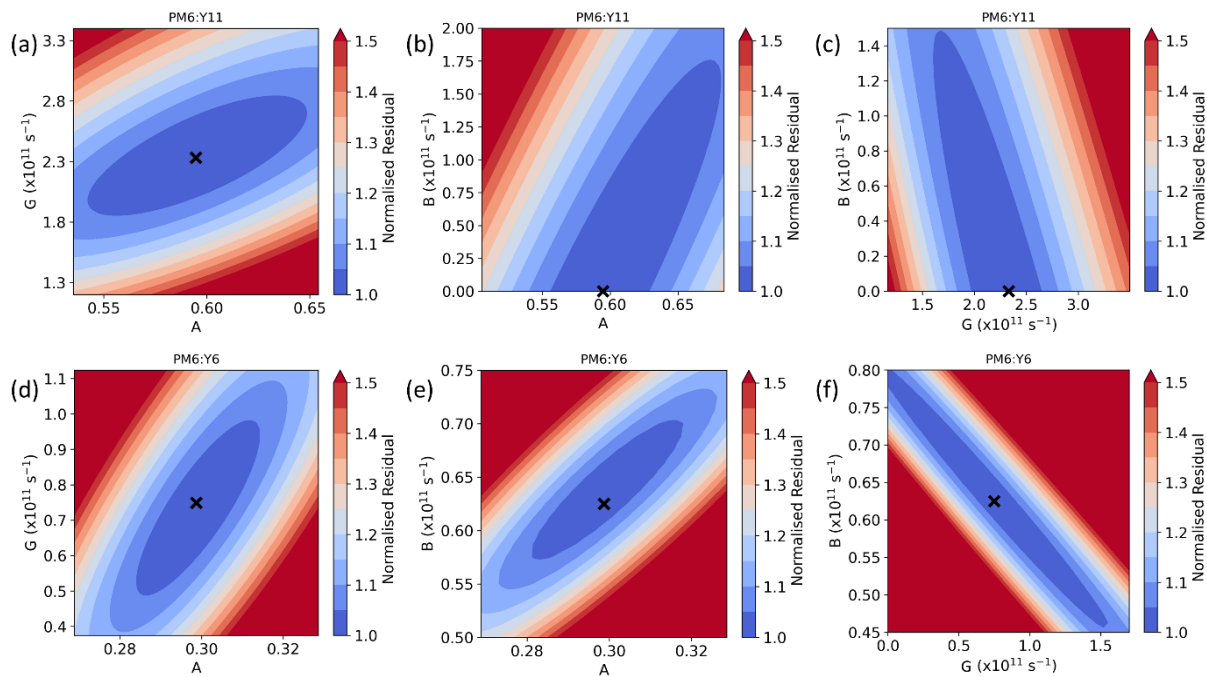

**Figure S7**: The normalised residuals obtained from fitting equation 3 to the femtosecond TAS data for (a-c) PM6:Y11 and (d-f) PM6:Y6. The colour indicates the magnitude of the residual normalised to its optimised value. For each blend, the magnitude of the residual has been plotted for all possible combinations of the fitting parameters *A*, *B* and *G*, which are directly proportional to α, β and γ, respectively (see S1.1). The black crosses indicate the optimal values of the parameters being varied. For PM6:Y11, the optimised value of *B* is negligible (figures b-c). We ascribe this behaviour to the high rate of TTA in PM6:Y11, which means that it outcompetes TCA as the dominant triplet decay channel (Figure S8). Thus, very few triplet excitons decay via TCA, preventing us from extracting a value for its rate constant. The low rate of TCA in PM6:Y11 may be due to its high rate of non-geminate charge recombination (see Figure S9), which leads to a comparatively small hole polaron PIA, as can be seen in Figure S22.

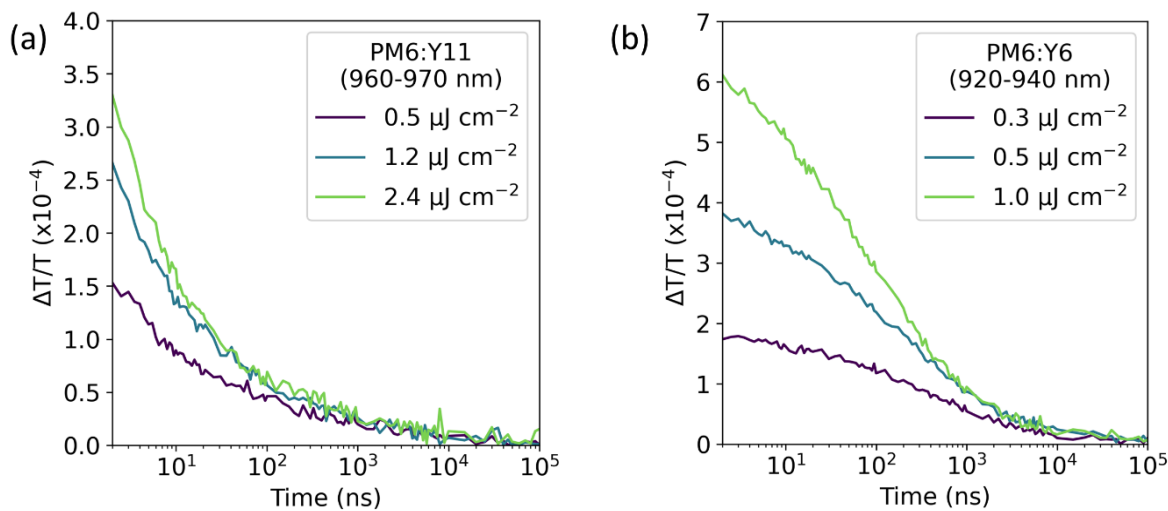

**Figure S9:** The nanosecond TAS data of (a) PM6:Y11 and (b) PM6:Y6 films in the PM6 hole polaron region following an excitation at 532 nm. It is clear that the decay of the PM6 hole polaron signal is accelerated in PM6:Y11 when compared to PM6:Y6, indicating a significantly higher rate of non-geminate charge recombination in the former blend, as is also observed in the femtosecond TAS data of the same wavelength region (Figure S22). Since the rate at which TCA occurs depends not only on its rate constant, β, but also on the population of charges, the low charge population in PM6:Y11 suppresses the effective rate of TCA and may contribute to the fact that a value for $B$ could not be extracted.

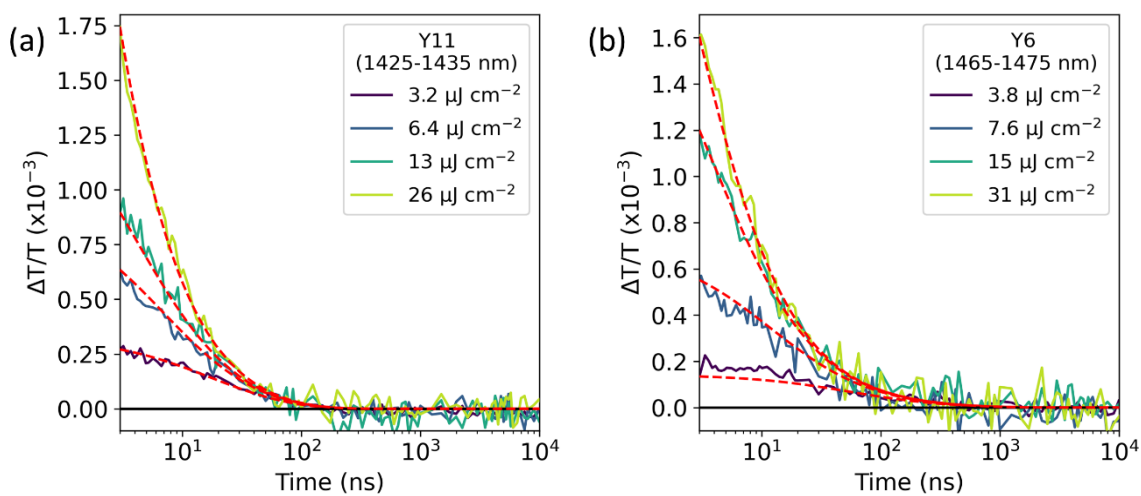

**Figure S10:** The nanosecond TAS data of (a) neat Y11 and (b) neat Y6 films in the triplet exciton region following excitation at 800 nm. The red dashed lines indicate the fitting results. The fitting methodology is described in S1.2, and the fitting parameters are given in Table S1. The decay of both signals is found to be well-modelled by a combination of TTA and monomolecular triplet decay. The triplet lifetimes extracted from the fits exceed the timespan of the femtosecond TAS measurements, justifying the exclusion of a monomolecular decay term from equation 3 in the main text.

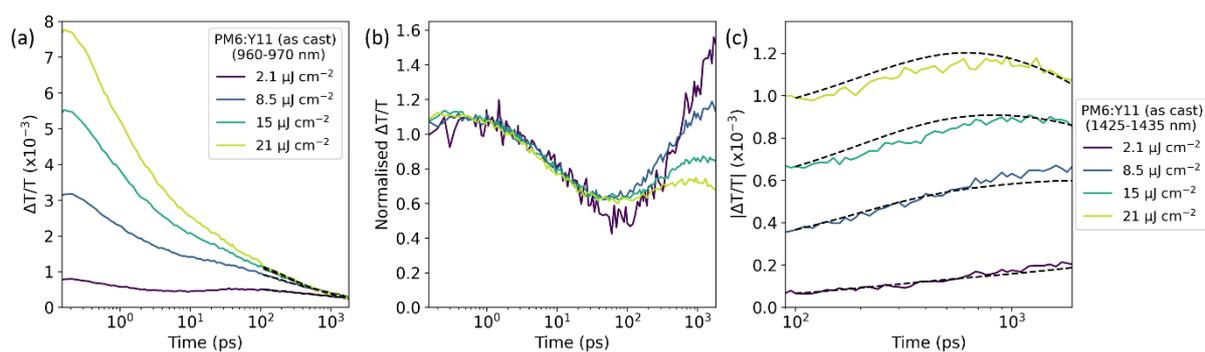

**Figure S11:** Femtosecond TAS data and fitting results for unannealed PM6:Y11 following excitation at 800 nm. (a) Fluence series of the hole polaron region. The black dotted lines indicate the fit obtained using a double exponential decay in order to extract an expression for the hole polaron signal and its derivative at each value of time. As was the case for the annealed PM6:Y11, the polaron signal is small at times >100 ps, meaning that a reliable TCA rate constant could not be extracted from the data. (b) The fluence series of the triplet region, normalised to the peak of the inter-CT PIA around t = 0. The fluence dependence is qualitatively the same as was observed in the annealed PM6:Y11, indicating that triplet decay is still dominated by TTA in the unannealed sample. (c) Results of the global fit for the Y11 triplet population. The fitting parameters were $A = 0.50 \pm 0.03$ and $G = (1.7 \pm 0.3) \times 10^{11}$ s$^{-1}$.

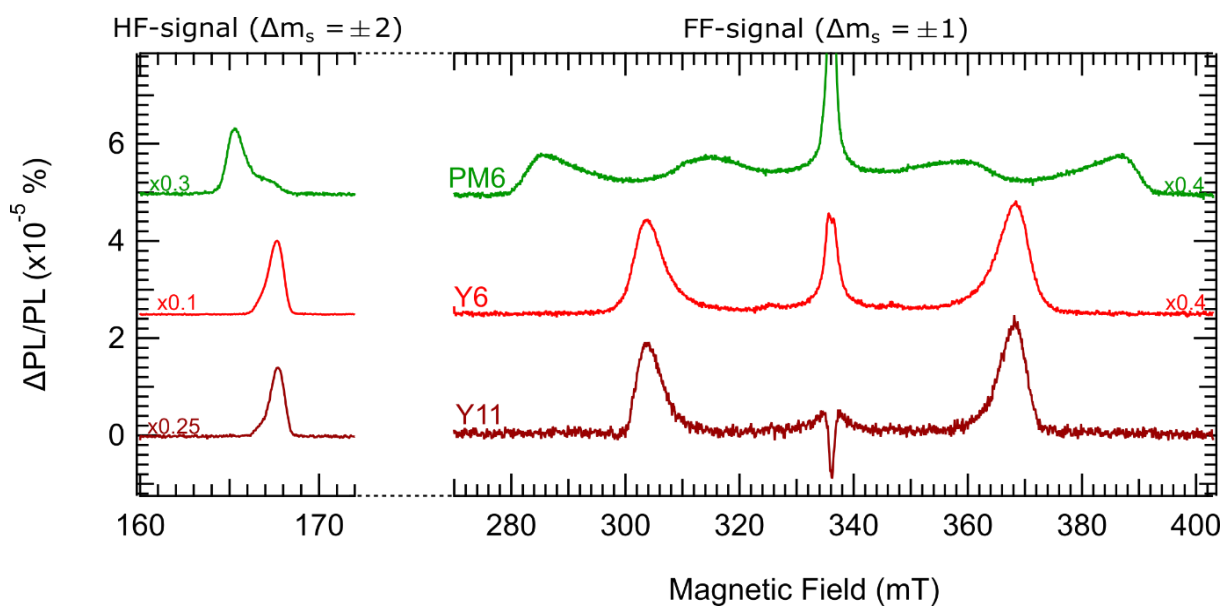

**Figure S12:** cwPLDMR for neat PM6, Y6 and Y11. The neat NFAs have similar spectral widths (correlated with the ZFS parameter *D*, see Table S2) and similar position of the half field (HF) signal, both of which are consistent with those measured in the blends PM6:Y6 and PM6:Y11. PM6 exhibits a larger ZFS splitting *D* (larger spectral width), also leading to a shift in the position of the HF signal.

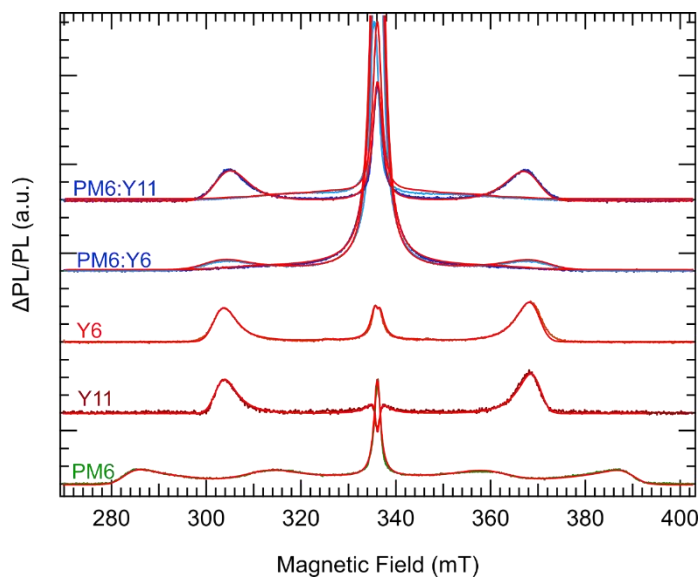

**Figure S13:** Spectral simulations of the cwPLDMR spectra produced using the MATLAB toolbox EasySpin and the parameters given in Table S2.

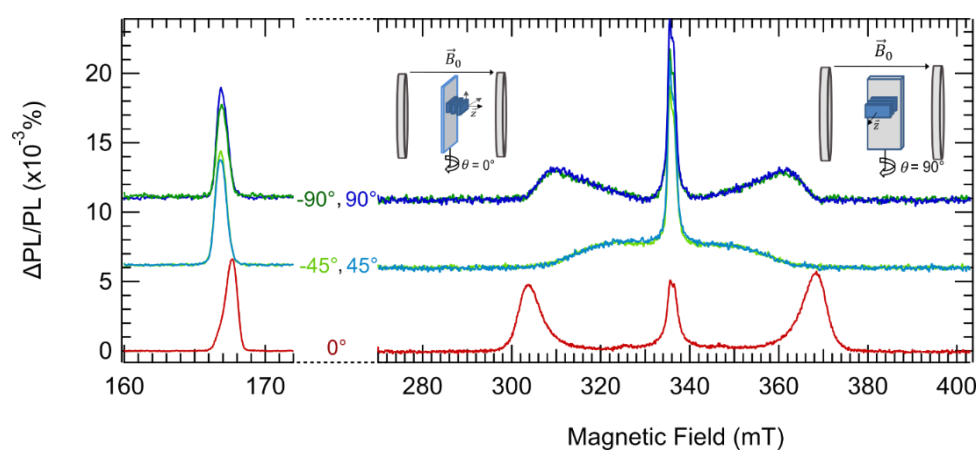

**Figure S14:** Rotation-dependent PLDMR spectra for neat Y6, discussed further in S2.2. The identical spectra measured at +45° and -45° confirm $C_{2v}$ symmetry, allowing the assignment of one principal ZFS axis which is perpendicular to the rotation axis. For $\theta = 0°$, the spectrum is determined by $D_z$ ($\vec{B}_0 \parallel D_z$), while at $\theta = 90°$, the spectrum is determined by $D_{x,y}$ components. Due to the structural similarity of Y6 and Y11, it is assumed that the same holds true for the latter molecule.

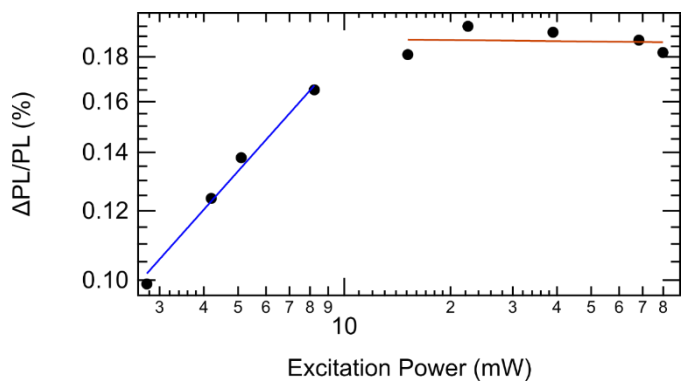

**Figure S15:** Laser excitation power dependence of relative PLDMR signal (ΔPL/PL) for PM6:Y11 from Figure 3a. The data points were fitted using the power law ΔPL/PL ~ $P_{exc}^a / P_{exc}^b = P_{exc}^c$ and two regimes of behaviour were identified. In the low-power regime, $c = 0.45 \pm 0.03$ and, in the high-power regime, $c < 0.03$. Considering the values of a given in Figure 3b, it follows that $b = 1$, i.e. the PL intensity depends linearly on the excitation power.

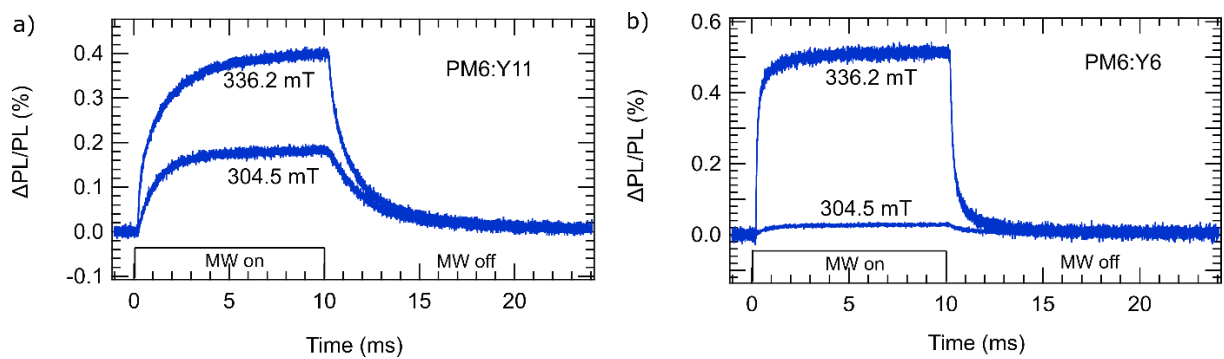

**Figure S16:** PLDMR transients for the middle peak ($B = 336.2$ mT) and the triplet feature ($B = 304.5$ mT). (a) For PM6:Y11, the ratio of the middle peak to the triplet feature in the steady state ($t = 9.7$ ms) is measured to be 0.45. (b) For PM6:Y6, the ratio of the middle peak to the triplet feature in steady state is measured to be 0.05, i.e. nine times smaller than that in PM6:Y11. Although the presence of TCA and inter-CT states can also enhance the middle peak and induce changes in signal shape at early times, the ninefold increase in the ratio of the middle peak to the triplet feature strongly suggests that triplet states in PM6:Y11 contribute more to the PL than those in PM6:Y6.

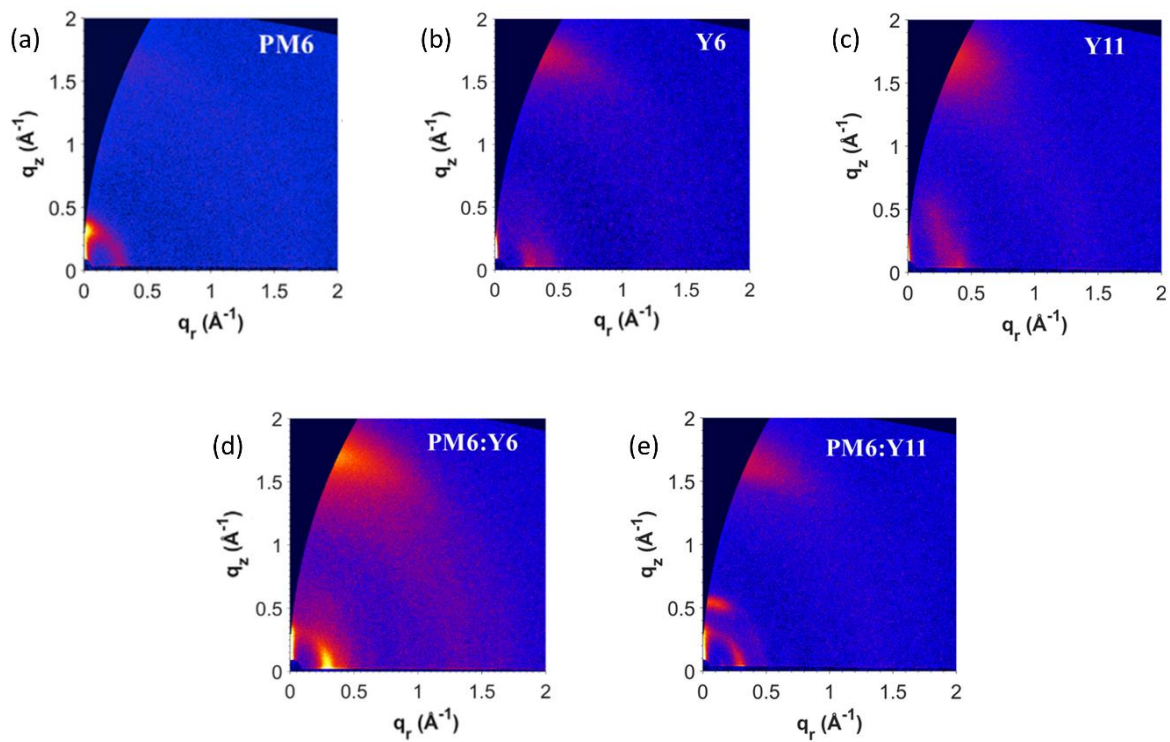

**Figure S17:** 2D GIWAXS images from which the line cuts in Figure 4 were taken. Reprinted with permission of Yuan et al [3].

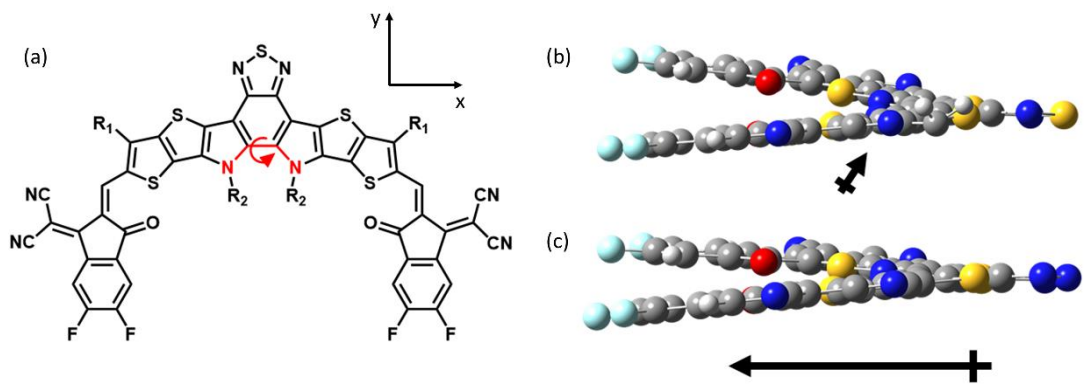

**Figure S18:** Density Functional Theory simulation results, with extracted parameters given in Table S4. (a) The molecular structure of Y6, on which the dihedral angle is highlighted in red. Both Y6 and Y11 are not planar due to the steric clash between the alkyl sidechains labelled $R_2$, which are the same in both molecules (Figure 1a). The axis indicates the directions of the $D_x$ and $D_y$ dipole moments, with the $D_z$ dipole moment pointing out of the page, through the exact $D_z$ orientation is highly influenced by the direction of the side chains. The minimised energy structures of (b) Y6 and (c) Y11 with the side chains removed for clarity. The dipoles' magnitude and direction are denoted by the arrows. Although Y6 and Y11 both have an A-DA'D-A structure, the central acceptor (A') groups differ between the two molecules with Y11 having benzotriazole (BTz) in the place of benzothiadiazole (BT). In Y6, the electron density around the central BT group is balanced by the peripheral A groups, resulting in a negligible dipole in the x-y plane. However, as BTz is less electron-withdrawing than BT [4,5], the dipole in Y11 is dominated by the peripheral A groups, resulting in an enhanced dipole in the x-y plane.

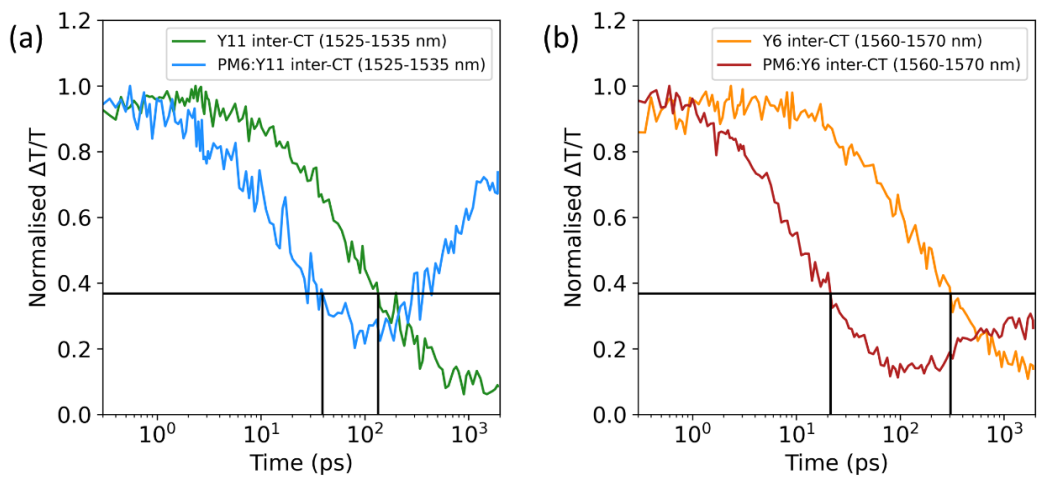

**Figure S19**: Reduction in the inter-CT state lifetime when moving from the neat NFA to the blend with PM6 for (a) PM6:Y11 and (b) PM6:Y6. Here, we use the low fluence measurement on the neat films due to the fluence dependence of the kinetics at early times, as commented upon in the caption of Figure S6. The reduction in the lifetime of the inter-CT state when we go from the neat film to the blend can be used to estimate the charge transfer efficiency ($\eta_{CT}$), as described in S3.1. For PM6:Y6, we calculate $\eta_{CT}$ = 0.93, whereas it is only 0.71 for PM6:Y11. Although this could indicate a lower yield of charges in PM6:Y11, considering that the peak EQE$_{PV}$ in the NFA spectral region for PM6:Y11 is about the same as PM6:Y6 (Figure S1), a more probable explanation is that significant charge generation occurs directly from the Y11 singlet exciton. This hypothesis is also supported by the lower yield of inter-CT states from singlet excitons observed in PM6:Y11, as is discussed in the caption of Figure S4.

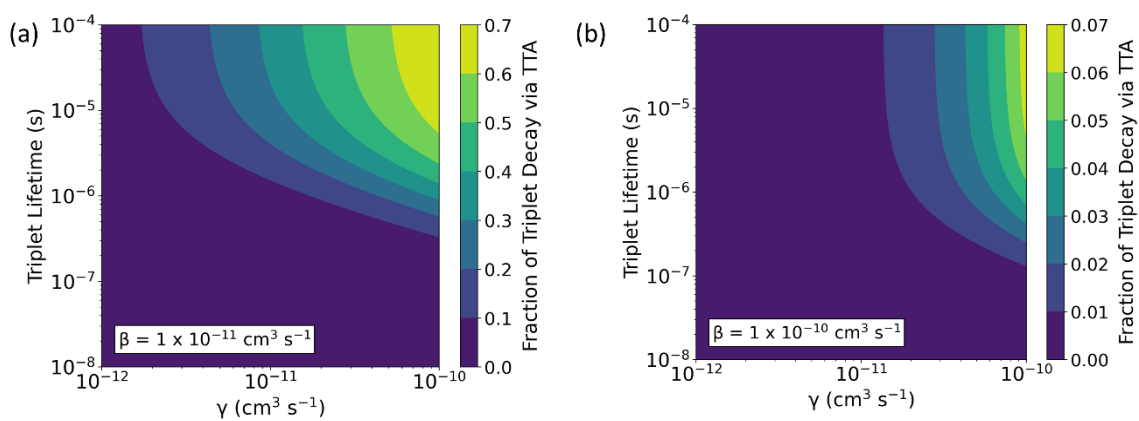

**Figure S20:** The fraction of the total triplet decay which occurs via TTA for TCA rate constants ($\beta$) of (a) 1 x 10$^{-11}$ cm$^3$ s$^{-1}$ and (b) 1 x 10$^{-10}$ cm$^3$s$^{-1}$. These higher rates of TCA are more representative of the situation in PM6:Y6 and demonstrate that TTA is unlikely to be a significant triple decay pathway under 1-Sun conditions, given the triplet lifetime of ~400 ns in this blend (see Table S1).

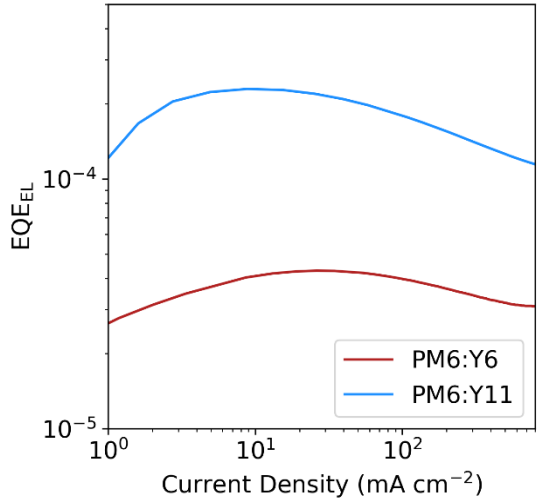

**Figure S21:** The electroluminescent external quantum efficiency (EQE$_{EL}$) of the PM6:Y6 and PM6:Y11 devices whose performance is shown in Figure S1. At an injected current density of 20 mA cm$^{-2}$, giving carrier densities approximately equivalent to those at short-circuit under 1-Sun conditions, the EQE$_{EL}$ of PM6:Y11 (2.2 x 10$^{-4}$) is five times higher than that of PM6:Y6 (4.3 x 10$^{-5}$). Using Rau's reciprocity relationship, this corresponds to a non-radiative voltage loss of 220 mV in PM6:Y11 and 260 mV in PM6:Y6.

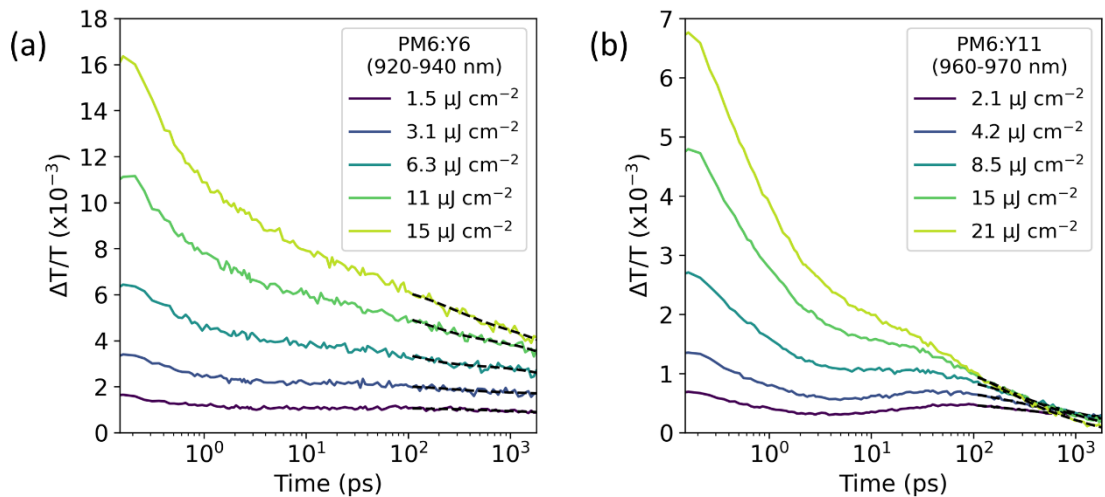

**Figure S22:** Fluence series of the hole polaron region in (a) PM6:Y6 and (b) PM6:Y11 following excitation at 800 nm. The black dotted lines indicate the fits to the data obtained using a double exponential decay, which were subsequently used to perform the global fit, as described in S1.1.

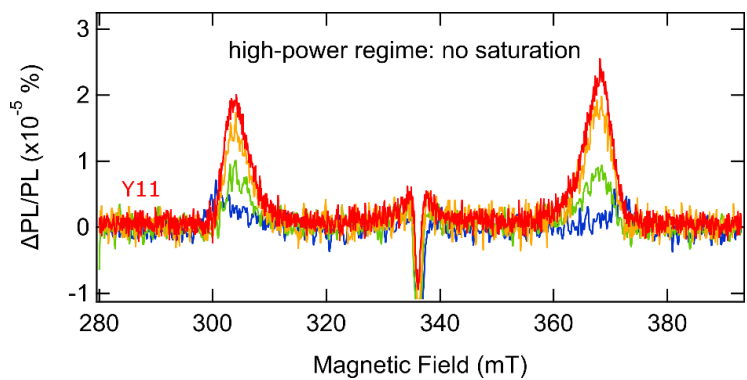

**Figure S23:** Power dependence of neat Y11's relative PLDMR signal in the high-power regime, discussed further in S2.3. In contrast to Figure 3b and Figure S14, the relative PLDMR signal does not plateau.

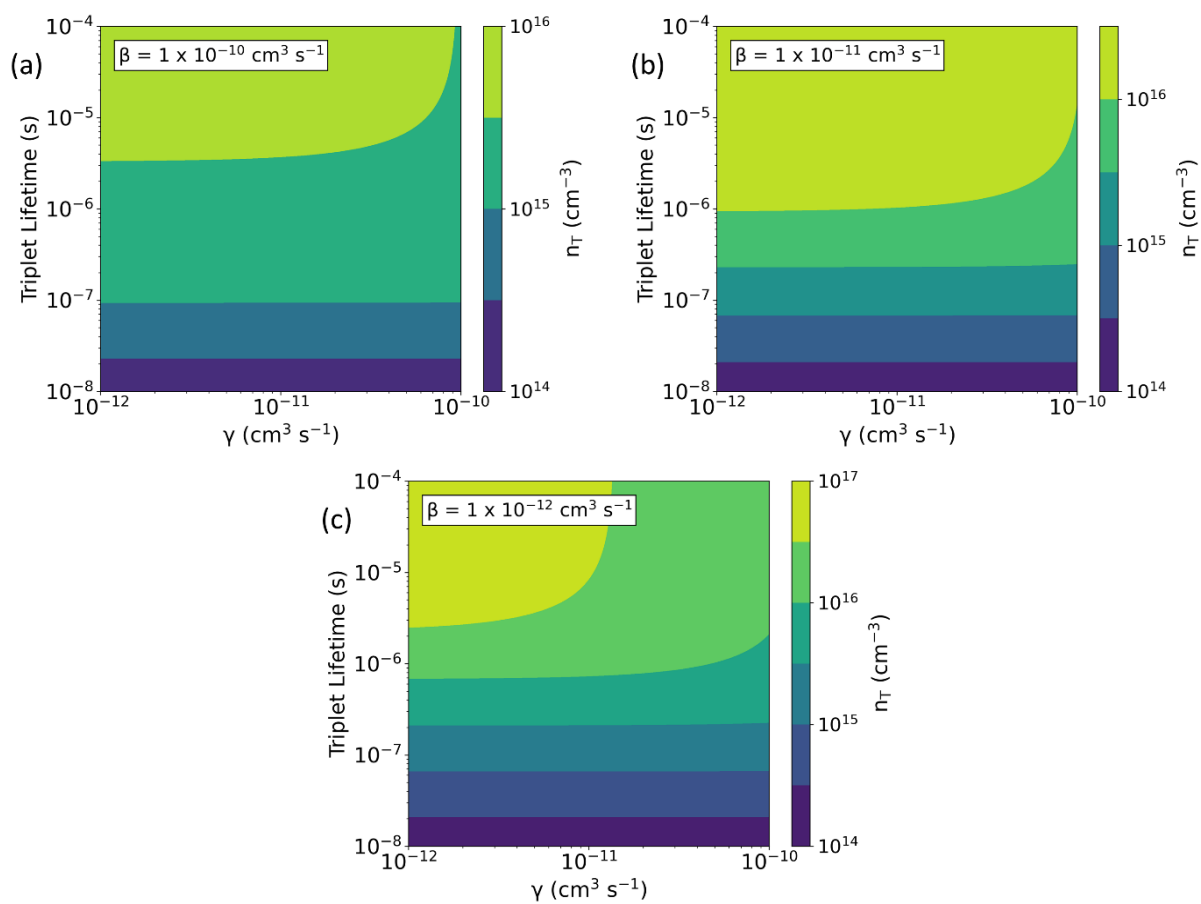

**Figure S24:** Triplet populations as a function of triplet lifetime and TTA rate ($\gamma$) for three different rates of TCA ($\beta$). The values were estimated by solving equation S7 under steady-state conditions, as described in S3.2. For blends with triplet lifetimes in the range 10-100 ns, as our fits to nanosecond TAS data indicate is the case for PM6:Y6 and PM6:Y11 (see Table S1), these plots indicate a steady-state triplet population of ~ 1 x $10^{15}$ cm$^{-3}$ under 1-Sun conditions.

# Supplementary Methods

## S1. Calculating Triplet Decay Rates from Transient Absorption Spectra

*S1.1 Rate Equations for Blend Films*

To model the triplet kinetics in the blend films, we used equation (3) in the main text, which is reproduced here for clarity:

$$\frac{dn_T}{dt} = -\alpha \frac{dn_H}{dt} - \beta n_P n_T - \gamma n_T^2 \tag{S1}$$

$n_T$ and $n_H$ refer to the triplet and hole population densities, respectively, α is the fraction of non-geminate recombination which leads to triplet formation, and β and γ are the rate constants of TCA and TTA. This equation can be linked to the measured values of ΔT/T via

$$\frac{\Delta T(\lambda, t)}{T} = -w\sigma(\lambda)\Delta n(\lambda, t) \tag{S2}$$

w is the film width, σ(λ) is the absorption cross-section and Δn(λ,t) is the density of the excited species (averaged over the film width). To fit equation S1 to the TAS spectra directly, equation S2 was used to transform equation S1 into the 'cross-section free' form given in equation 4 of the main text. As part of this process, α, β and γ were transformed into new parameters (*A*, *B* and *G*) which have no dependence on either the absorption cross sections or the film thickness [6]. The parameters are related to one another by the transformations:

$$A = \frac{\sigma_T}{\sigma_P}\alpha \; ; \; B = \frac{\beta}{w\sigma_P} \; ; \; G = \frac{\gamma}{w\sigma_T} \tag{S3}$$

$\sigma_P$ is the polaron cross-section and $\sigma_T$ is the triplet cross-section. We then performed a least-squares fit of equation 4 to the triplet region TAS spectra (shown in Figures 2b and 2e) using the Python package, *LMFIT* (v 1.0.3 [7]). To perform this fit, the hole polaron region TA spectra (shown in Figure S22) were modelled using a bi-exponential function so that the value of ΔT/T and its derivative in this spectral region could be found each time point. From this process, we obtained the optimal values of *A*, *B* and *G* (see Table 1 in the main text), which were each assumed to be independent of fluence (i.e., a global fit). For *B* and *G*, this assumption is valid as the low levels of energetic disorder in these blends means that second order non-geminate rate constants are not expected to be fluence dependent [8,9]. *A* is also assumed to be fluence independent as there is no obvious mechanism by which the fluence could affect the fraction of non-geminate recombination which forms triplet states.

As commented upon in the main text, it was not possible to extract a value of *B* for PM6:Y11. Instead, the optimal value of *B* was found to be negligible, which is unphysical as TCA will still be a possible triplet decay mechanism in PM6:Y11. Instead, this indicates that the rate of TCA in PM6:Y11 is sufficiently slow when compared to the rate of TTA that TTA is the dominant triplet recombination pathway at all the fluences probed in this study and thus a rate constant for TCA cannot be extracted from the data. This conclusion is also supported by Figure S8 where we demonstrate that, even if the rate constant in PM6:Y11 takes the same value as in PM6:Y6, TTA still dominates triplet decay. The low apparent rate of TCA in PM6:Y11 may be due to its high rate of non-geminate recombination (Figure S9), which leads to a relatively weak hole polaron PIA (Figure S22) as charges rapidly recombine with one another to form CT states, rather than remaining in the blend for long enough to undergo TCA. The high rate of non-geminate recombination in PM6:Y11 is surprising given the reasonable efficiency of the PM6:Y11 device (Figure S1a) and suggests that the enhanced crystallinity of the Y11 domains increases the rate constants of all recombination pathways, not just TTA.

*S1.2 Rate Equations for Neat Y-Series Films*

In the neat NFA films at times greater than 2 ns, we assume that all the excited states except for triplet excitons generated by inter-system crossing (ISC) have returned to the ground state (Figure S6). Thus, there will be no further triplet generation and triplets will be unable to decay via TCA. However, due to the relatively low density of triplet excitons generated by ISC, monomolecular triplet decay will compete with TTA to be the dominant triplet decay mechanism, especially at low fluences. This means that the rate equation for the triplet exciton population in the neat NFA films takes the form

$$\frac{dn_T}{dt} = -k n_T - \gamma n_T^2 \tag{S4}$$

where k is the rate of monomolecular triplet decay, which is the reciprocal of the monomolecular triplet lifetime. As for equation S1, equation S4 was transformed to replace γ with the cross-section free parameter, $G$ (Equation S3). The transformed version of equation S4 has the analytic solution

$$\Delta T_T(t) = \frac{k}{2G}\left[\frac{1+C\exp(-kt)}{1-C\exp(-kt)} - 1\right]; \quad C = \left(\frac{\Delta T_{T,0}}{\Delta T_{T,0}+k/G}\right)\exp(kt_0) \tag{S5}$$

$\Delta T_T(t)$ is the measured value of $\Delta T/T$ at time t, $t_0$ is the time from which the fitting begins and $\Delta T_{T,0}$ is the value of $\Delta T/T$ at time $t_0$. We note that this solution allows us to extract a value for the triplet lifetime, τ, from the fit results without any knowledge of the absorption cross-section or film width by defining τ = 1/k. Equation S5 was fitted globally to the nanosecond TAS data, as is shown in Figure S10, and the fitting parameters are given in Table S1. The value of the triplet lifetime extracted using this method has a high uncertainty for the neat Y6 film, which may be due to the low signal to noise ratio at values of t > 100 ns. Despite this, the τ values for the neat NFA films are both significantly longer than the timescale of the femtosecond TAS measurements, which extend up to 2 ns. Thus, we are justified in excluding a term describing monomolecular triplet decay from equation 3, as it would not have a significant impact on the triplet kinetics on such short timescales.

## S2. Photoluminescent Dependent Magnetic Resonance (PLDMR)

*S2.1 Principles of PLDMR*

PLDMR probes the relative change of PL under resonant conditions, i.e., when the energy of the applied microwave irradiation corresponds to the splitting of triplet sublevels induced by the Zeeman and dipolar interactions. Microwave irradiation alters the net spin polarisation of the sample by inducing transitions between triplet sublevels, resulting in a change of the overall PL. The width of the full-field (FF) spectrum is determined by the axial ZFS parameter $D$ by $|2D|\hbar/g\mu_B$ whereby $g$ represents the g-factor (g-tensor assumed to be isotropic due to small spin-orbit coupling in organic molecules) and $\mu_B$ the Bohr magneton. In organic materials, the parameter $D$ is mainly determined by dipolar interactions and thus depends on the delocalization, $r$, of the paramagnetic spin species [10,11]. The ZFS parameter $E$ is a measure of the rhombicity and thus of the deviation from axial symmetry [11,12]. While CT states possess a small dipolar interaction, nearby spins in molecular triplet excitons possess a considerable $D$ value, allowing the spectral width to be used as an indicator for their molecular assignment. An additional feature is the half-field (HF) signal, corresponding to the first-order forbidden $\Delta m_S = \pm 2$ transition between $T_+$ and $T_-$ sublevels. The probability of this transition increases with $I_{HF} \sim D^2 \sim r^{-6}$, leading to a higher signal intensity for close-by spins, i.e., predominantly visible for molecular triplet excitons [13]. As the position of the HF signal depends on the ZFS parameters, it is an additional tool for determining the molecular affiliation of the probed triplet states [13]. Thus, comparing the HF signals and the $D$ values of the FF spectrum of blends and the neat materials (Figure 3a and Figure S12), the HF signal confirms the detection of Y11 and Y6 triplet excitons in the blends, in agreement with TAS findings.

*S2.2 Rotational Dependence of PLDMR*

The transitions in the PLDMR spectrum depend on the orientation of the principal axes $X$, $Y$ and $Z$ of the ZFS tensor with respect to the external magnetic field $\vec{B}_0$. In the so-called canonical orientations, the external magnetic field is aligned with one of the principal axes, while the different energetic splitting in high field leads to EPR transitions at different magnetic field values [14]. The spectral separation between the transitions is proportional to the energetic splitting of |2D|, |D|+3|E| and |D|-3|E| for $Z||\vec{B}_0$, $Y||\vec{B}_0$ and $X||\vec{B}_0$, respectively [11,12,14]. In the PLDMR spectrum, transitions from all orientations of the ZFS tensor with respect to the external magnetic field $\vec{B}_0$ are superimposed. If the sample is disordered, i.e. randomly oriented molecules, each orientation of the ZFS tensor occurs with equal probability, leading to a superimposed spectrum that is independent on the angle to the magnetic field [17]. However, if certain orientations are more probable than others, the system is partially ordered, whereby the anisotropic orientational distribution can be weighted by an ordering parameter, based on the averaged second Legendre polynomial $\langle P_2(\cos\theta) \rangle = \langle \frac{1}{2}(3\cos^2\theta - 1) \rangle$ [15,16].

Figure S14 shows the PLDMR spectra of neat Y6 between -90 and 90° in 45° steps. While the identical spectra for -90 and 90° degree reveal $C_2$ symmetry, the identical spectra for -45° and 45° reveal $C_{2v}$ symmetry, consistent with the symmetry of neat Y6 [16]. This symmetry implies that the rotation axis is perpendicular to one of the principal axes of the ZFS tensor [16]. At 0°, the Z-transitions dominate the PLDMR spectrum, and so its width is determined by the $D$ value, while for -90° and 90°, X- and Y-transitions are visible, the width of which are determined by the $E$ value. Given this orientation dependence, we can conclude that the $D_z$ component is parallel to the external magnetic field at 0°. GIWAXS measurements show intermolecular face-on stacking between Y6 molecules and face-on stacking on the substrate. For 0°, the OOP direction of the face-on stacking is parallel to the external magnetic field (right inset Figure S14). Thus, at 0°, the $D_z$ component and OOP direction are parallel to one another, meaning that the pronounced Z-transitions are an indicator of a preferential orientation or stacking in z-direction, i.e., OOP-direction. Thus, the steeper wings in PM6:Y11 (Figure 3a) indicate a stronger alignment of the paramagnetic molecules at 0°, i.e., higher crystallinity in the OOP direction.

*S2.3 Pump Intensity Dependent PLDMR Spectra of neat Y11*

Figure S23 shows the excitation power dependent PLDMR signals of neat Y11, where triplet excitons predominantly stem from ISC (Figure S6). Unlike excitation power dependent measurements for PM6:Y11 (Figure 3b), the ΔPL/PL signals for neat Y11 do not plateau at high powers. When Y11 is blended with PM6, the high rate of charge transfer reduces the proportion of excitons which persist long enough to undergo ISC. However, the generation of triplet excitons by non-geminate recombination significantly increases the triplet population in PM6:Y11 when compared to neat Y11, leading to the presence of the annihilation-limited regime at lower fluences in the blend film.

## S3. Calculation of the Reduction in $\Delta V_{nr}$ due to TTA

*S3.1 Estimating the Charge-Transfer Efficiency*

To calculate the improvement in $EQE_{EL}$ due to TTA, it is necessary to estimate the fraction of singlet excitons which dissociate to form charges ($\eta_{CT}$). To do this, we use the fact that the majority of singlet excitons form inter-CT states on Y6, prior to charge separation [2]. Thus, by measuring the reduction in the lifetime of the inter-CT state PIA when moving from neat Y6 to the PM6:Y6, we can estimate the efficiency of charge transfer from the inter-CT state. Specifically, we calculate the quantity

$$1 - \frac{\tau_{PM6:Y6}}{\tau_{Y6}} = \frac{k_{CT}}{k_{CT} + \frac{1}{\tau_{Y6}}} \tag{S6}$$

$k_{CT}$ is the rate of charge transfer, $\tau_{Y6}$ the 1/e time of the inter-CT PIA in the neat Y6 film and $\tau_{PM6:Y6}$ the 1/e time of the inter-CT PIA in the PM6:Y6 film. This is illustrated in Figure S19 where, for PM6:Y6, we find that $\tau_{Y6} \sim$ 300 ps and $\tau_{PM6:Y6} \sim$ 20 ps, giving $\eta_{CT}$ = 0.93. This value agrees well with the high values of IQE which have been reported for PM6:Y6 previously [17,18].

We note that this method of estimating $\eta_{CT}$ may not be valid for PM6:Y11. As is shown in Figure S19, the value of $\eta_{CT}$ calculated for PM6:Y11 is only 0.71 if we assume that all charges are generated from the inter-CT state. This would imply that charge generation is ~25% less efficient in PM6:Y11 than PM6:Y6. However, this cannot be the case as the EQE$_{PV}$ values of PM6:Y11 and PM6:Y6 are comparable at 800 nm (Figure S1b). Thus, it is more likely that there is significant charge generation directly from the Y11 singlet state, meaning that $\eta_{CT}$ cannot be estimated from the reduction in the lifetime of the inter-CT state in the PM6:Y11 blend.

*S3.2 Calculation of the Fraction of Triplet Decay via Triplet-Triplet Annihilation*

To calculate the fraction of triplet decay which occurs via TTA, we must estimate the triplet population density under 1-Sun illumination as a function of the triplet lifetime ($\tau$), TCA rate constant ($\beta$) and TTA rate constant ($\gamma$). To do this, we solve the rate equation

$$\frac{dn_T}{dt} = \alpha k_2 n_P^2 - \beta n_P n_T - \gamma n_T^2 - \frac{1}{\tau} n_T \tag{S7}$$

under steady-state conditions i.e., where the left-hand side is equal to zero. In equation S7, $n_T$ and $n_P$ refer to the triplet and hole polaron population densities, respectively, α is the fraction of non-geminate recombination which leads to triplet formation and $k_2$ is the rate of the bi-molecular recombination of electron and hole polarons. To find a value for the triplet population under 1-Sun conditions, we must assume values for α, $k_2$ and the hole polaron density. For the latter two quantities, we have taken the values in ref. 9 as representative for those reported for PM6:Y6 in the literature ($n_P = 4.5 \times 10^{16}$ cm$^{-3}$ and $k_2 = 1 \times 10^{-11}$ cm$^3$ s$^{-1}$). However, we note that these values can vary between active layer blends which use different Y-series NFAs [19] and so these values may not be representative of PM6:Y11, especially if this blend does have a higher rate of non-geminate charge recombination, as suggested by Figures S9 and S22. We have assumed that the value of α is 0.75, in line with the naïve prediction of spin-statistics, and we did not find that there was a significant impact on our conclusions for values of α in the range 0.5-1.0. The values of $n_T$ calculated using this method are shown in Figure S24 as a function of $\gamma$ and $\tau$ for three values of $\beta$. We note that, for a triplet lifetime in the range of 10-100s of nanoseconds, as found for Y6 and Y11, the triplet population at steady-state is always lower than the assumed steady-state polaron population. This justifies our assumption in the main text that the polaron population is larger than the triplet population, and thus that TCA is the dominant triplet decay channel in PM6:Y6 (as is also illustrated in Figure S8).

Once the value of $n_T$ is known, each of the three triplet decay terms in equation S7 can be quantified and the fraction of the total triplet decay occurring via TTA, *f*, can be calculated via

$$f = \frac{\gamma n_T^2}{\frac{1}{\tau} n_T + \beta n_P n_T + \gamma n_T^2} \tag{S8}$$

**Supplementary References**